\documentclass[prd,twocolumn,nopacs,nofootinbib,preprintnumbers]{revtex4}

\usepackage{caption}
\usepackage{multirow}
\usepackage{mathrsfs}
\usepackage{yfonts}
\usepackage{tipa}
\usepackage{bm}
\usepackage{bbm}
\usepackage{amsmath}
\usepackage{amssymb}
\usepackage{amsthm}
\usepackage{amsfonts}
\usepackage{mathtools}
\usepackage{color}
\usepackage{latexsym}
\usepackage{relsize}
\usepackage[greek,english]{babel}
\usepackage{booktabs}
\usepackage{hyperref}

\usepackage{array}
\newcolumntype{?}{!{\vrule width 2pt}}

\newcommand{\lp}{\left(}
\newcommand{\rp}{\right)}
\newcommand{\lb}{\left[}
\newcommand{\rb}{\right]}

\newcommand{\T}{\mathfrak{T}}

\newcommand{\mS}{\mathcal{S}}
\newcommand{\mQ}{\mathcal{Q}}

\newcommand{\hL}{\hat{ L}}

\newcommand{\mL}{{\mathcal L}}

\newcommand{\lsim}   {\mathrel{\mathop{\kern 0pt \rlap
  {\raise.2ex\hbox{$<$}}}
  \lower.9ex\hbox{\kern-.190em $\sim$}}}
\newcommand{\gsim}   {\mathrel{\mathop{\kern 0pt \rlap
  {\raise.2ex\hbox{$>$}}}
  \lower.9ex\hbox{\kern-.190em $\sim$}}}
\newcommand{\bw}{\begin{widetext}\begin{equation}}
\newcommand{\ew}{\end{equation}\end{widetext}}
\newcommand{\be}{\begin{equation}}
\newcommand{\ee}{\end{equation}}
\newcommand{\ba}{\begin{eqnarray}}
\newcommand{\ea}{\end{eqnarray}}

\newcommand{\diff}{{{\rm d}}}

\newcommand{\nn}{\nonumber}

\begin{document}

\title{Teleparallel Palatini theories}

\author{Jose Beltr\'an Jim\'enez}
\email{jose.beltran@uam.es}
\affiliation{Instituto de F\'isica Te\'orica UAM-CSIC, Universidad Aut\'onoma de Madrid, Cantoblanco, Madrid, 28049 Spain\\
Departamento de F\'isica Fundamental, Universidad de Salamanca, E-37008 Salamanca, Spain.}

\author{Lavinia Heisenberg}
\email{lavinia.heisenberg@eth-its.ethz.ch}
\affiliation{Institute for Theoretical Studies, ETH Zurich, Clausiusstrasse 47, 8092 Zurich, Switzerland}

\author{Tomi S.  Koivisto}
\email{tomik@astro.uio.no}
\affiliation{Nordita, KTH Royal Institute of Technology and Stockholm University, Roslagstullsbacken 23, 10691 Stockholm, Sweden}
\affiliation{Helsinki Institute of Physics, P.O. Box 64, FIN-00014 Helsinki, Finland} 
\affiliation{Department of Physical Sciences, Helsinki University, P.O. Box 64, FIN-00014 Helsinki, Finland}

\preprint{IFT-UAM/CSIC-18-035}
\preprint{NORDITA-2018-023}

\date{\today}

\begin{abstract}

The Palatini formalism, which assumes the metric and the affine connection as independent variables, is developed for gravitational theories in flat geometries. We focus on two particularly interesting scenarios. First, we fix the connection to be metric compatible, as done in the usual teleparallel theories, but we follow a completely covariant approach by imposing the constraints with suitable Lagrange multipliers. For a general quadratic theory we show how torsion naturally propagates and we reproduce the Teleparallel Equivalent of General Relativity as a particular quadratic action that features an additional Lorentz symmetry. We then study the much less explored theories formulated in a geometry with neither curvature nor torsion, so that all the geometrical information is encoded in the non-metricity. We discuss how this geometrical framework leads to a purely inertial connection that can thus be completely removed by a coordinate gauge choice, the coincident gauge. From the quadratic theory we recover a simpler formulation of General Relativity in the form of the Einstein action, which enjoys an enhanced symmetry that reduces to a second linearised diffeomorphism at linear order. More general theories in both geometries can be formulated consistently by taking into account the inertial connection and the associated additional degrees of freedom. As immediate applications, the new cosmological equations and their Newtonian limit are considered, where the role of the lapse in the consistency of the equations is clarified, and the Schwarzschild black hole entropy is computed by evaluating the corresponding Euclidean action. We discuss how the boundary terms in the usual formulation of General Relativity are related to different choices of coordinates in its coincident version and show that in isotropic coordinates the Euclidean action is finite without the need to introduce boundary or normalisation terms. Finally, we 
discuss the double-copy structure of the gravity amplitudes and the bootstrapping of gravity within the framework of coincident General Relativity.

\end{abstract}

\maketitle

\section{Introduction}
\label{introduction}

Special Relativity can be understood as the global, and General Relativity (GR) as the local pseudo-rotational symmetry in 1+3 dimensions. 
Poincar\'e gauge theory i.e. the theory of the inhomogeneous Lorentz group, is known long since to underpin GR \cite{Kibble:1961ba,Hehl:1976kj}. 
By a suitable prescription of the gauge geometry, the GR dynamics can be equivalently described either in a torsion-free but curved spacetime, or 
in a flat but contorted spacetime\footnote{For the sake of clarity, let us specify that a flat spacetime has vanishing total curvature, while contorted refers to the torsion piece of the connection, see Eq.(\ref{decomposition}). The same object is often called ''contorsion''.}. The former corresponds to Einstein's original formulation of GR and, a few years after this inception, Einstein himself attempted a unification \cite{Goenner2004} in the latter kind of spacetime, dubbed {\it teleparallel} due to a well-defined notion of parallelism owed to the absence of curvature \cite{PELLEGRINI:19165,Hayashi:1979qx,Mielke:1992te}, which yet today remains an efficient laboratory in the research into the remaining problems in gravity \cite{Baez:2012bn,Hohmann:2015pva,Hohmann:2017duq}.
 
It is then pertinent to begin by substantiating what might be gained by a mere change of the geometric stage (see \cite{Combi:2017crv} for an axiomatic
and \cite{Hehl:2016glb,Pfeifer:2016har} for pre-metric recent discussions).
While it has been realised that in the Teleparallel Equivalent of GR (TEGR) \cite{Aldrovandi:2013wha,Maluf:2013gaa}, there exists a tensor that defines the gravitational stress energy \cite{moller,deAndrade:2000kr,Maluf:2002zc,Maluf:2005kn}\footnote{This does not get around the Weinberg-Witten obstruction though \cite{Weinberg:1980kq}.}, the holographic and regularising roles of the teleparallel spin connection played at the boundary of the action integral have begun to be uncovered only 
recently\footnote{There is also cosmological interest in the dynamical role of the boundary terms in teleparallel modified gravity models \cite{Bahamonde:2015zma,Bahamonde:2016kba, 
delaCruz-Dombriz:2017lvj,Bahamonde:2016grb,Oshita:2017nhn}.}\cite{Lucas:2009nq,Krssak:2015rqa,Krssak:2015lba}, and the covariant definition of gravitational entropy\footnote{The thermodynamics of
torsional spacetime is considered in Ref. \cite{Dey:2017fld}, though not in teleparallelism.} remains to be clarified. Now, we can recall that in the Einstein-Hilbert action describing GR, the variational problem is not well-posed as it requires to impose the variations of both the metric and its normal derivatives to vanish upon the boundary of the integrated volume, or, otherwise we have to invoke a Gibbons-Hawking-York (GHY) boundary term that suitably cancels the higher derivatives from the variation. In a teleparallel formulation, the elegance and covariance of the theory needs not to be compromised, as the higher derivatives are absent in the action from the very beginning. The boundary term appearing in the teleparallel formulation represents
inertial effects, which typically diverge far away from the source. By the correct choice of the connection, the divergent inertial effects are removed 
locally at each point of spacetime, in contrast to the GHY normalisation that subtracts only their total integral from the action \cite{Lucas:2009nq,Krssak:2015rqa,Krssak:2015lba}. 

In TEGR \cite{Aldrovandi:2013wha,Maluf:2013gaa}, the spin connection enters the action as merely a surface term and is thus dynamically irrelevant.
When considering theories that modify TEGR, the spin connection does not cancel identically from the equations of motion, and one generically expects additional degrees of freedom to propagate. The covariant treatment of the theory requires to choose the tetrad according to the spin connection (or vice versa), and the latter can be erased only in a coordinate system that is suitably adopted for the former. The covariant formulation of the generalised teleparallel theories has been discussed recently
in the new terms of a reference tetrad that defines the spin connection \cite{Krssak:2017nlv}, and it was pointed out that already in the variational problem the teleparallel  
spin connection can  be taken to be of the inertial form \cite{Golovnev:2017dox}. Usually, the connection is also taken to be metric compatible, i.e. antisymmetric, a condition that could as well be alternatively imposed by a Lagrange multiplier in a wider geometric setting of asymmetric connections \cite{Hehl:1994ue,Obukhov:2002tm}. Recently, it was also reminded that by imposing the connection to be inertial with a Lagrange multiplier that sets the curvature to zero (in the metric compatible case), the physical content of the theory remains the same as in the non-covariant formulation, where one
just starts with a vanishing connection \cite{Nester:2017wau}, and this result can be slightly generalised to the minimal covariant formulation which accommodates an inertial 
(but not necessarily vanishing) connection \cite{Golovnev:2017dox}. Though the generalised models remain consistent with Lorentz covariance, the nature of the 
additional degrees of freedom has not been fully understood. Current investigations \cite{Ong:2013qja,Ferraro:2014owa,Chen:2014qtl,Nester:2017wau} have raised concerns about possible
problems such as acausalities, which might compromise especially the nonlinearly modified teleparallel gravity models that are often studied in cosmological applications \cite{Cai:2015emx,Hohmann:2017jao,Bahamonde:2015zma,Bahamonde:2016kba, 
delaCruz-Dombriz:2017lvj,Bahamonde:2016grb,Oshita:2017nhn}.  

In this article, we present the teleparallel geometries in the {\it Palatini formalism}. There, the independent gravitational variables are the metric and the connection;
for reviews on the Palatini formalism applied to various generalised gravitational theories - previously without the teleparallel restriction\footnote{Related to the reference tetrad approach to
teleparallel theory \cite{Krssak:2017nlv}, it has been considered that the affine connection is generated by an independent metric \cite{Koivisto:2011vq}. Regarding this independent metric as
a variational degree of freedom leads to what was called the bimetric variational principle \cite{BeltranJimenez:2012sz}, which, however, generates theories that are ghost-free only in very special cases
\cite{Golovnev:2014aca}. It could be interesting to check whether the case would be the same in a ''bitetrad variational principle'' that promoted the reference tetrad discussed in Ref.  \cite{Krssak:2017nlv} into a dynamical variable, and whether the bimetric interaction terms could be accommodated in such a theory. However, in this article we restrict to the standard Palatini formalism. \label{bitetrad}} 
 - see \cite{Olmo:2011uz,Capozziello:2015lza,BeltranJimenez:2017doy}. Note, that we avoid introducing the basic object of teleparallelism, a frame field. In the standard gauge formulation of gravity,
one sets up a principal bundle, with the horizontal base being the spacetime, the vertical transformations along the fibres corresponding to the homogeneous symmetries, 
and the geometry of the spacetime then realised through the projection by the frame field.  At each location of the spacetime, one can associate a vector space, the frame bundle, provided with a flat metric where the matter fields are connected by the homogeneous (spin) connection. The existence of a soldering form then allows to identify the frame and the tangent bundles. The fundamental objects are then, besides the frame field, the matter fields and their connection, since the action is extremised with respect to their variations, and from their solutions one derives the spacetime appearance of the relevant objects, in particular the metric $g_{\mu\nu}$ and affine connection $\Gamma^\alpha{}_{\mu\nu}$.
In the Palatini formalism\footnote{\label{footnote}This is common terminology though such first order formalism for spacetime geometry may have been introduced by A. Einstein \cite{turin,Goenner2004} independently of A. Palatini. Even though sometimes the first order variation in terms of tetrad and spin connection is also referred to as Palatini formalism, we mean by that, as usual in the context of modified gravity and cosmology \cite{Olmo:2011uz,Capozziello:2015lza,BeltranJimenez:2017doy}, strictly the variation wrt the spacetime metric and the spacetime affine connection. Further, we do not take the latter to be symmetric, which is sometimes (implicitly) assumed for the independent affine connection of the Palatini formalism.} we can deal with the generalised spacetime geometry more straightforwardly. We vary directly the spacetime metric and the spacetime connection so that there is no need to introduce the extra structure of a frame bundle and, consequently, the extra set of indices that would come with it. This might help to shed light on the open problems in (teleparallel) gravity we mentioned above. At least, no issues with covariance arise, since the actions constructed straight from spacetime tensors are trivially scalars once their indices are contracted away .

In teleparallel geometry, torsion can be interpreted as the external curvature, to wit it is the gauge field strength of the potentials that are dual to those generators of the gauge group that have been divided into the quotient that spans the tangent space of the base manifold. The torsion appearing, in particular, as the displacement field strength, the ensuing geometry devoid of homogeneous curvature indeed exhibits teleparallelism, parallelism at distances. However, since pure displacements can be realised by a series of disclinations, i.e. translations are obtained by rotations in higher dimensions, one is led to entertain the geometry where the curvature disappears for the total gauge group, i.e. also in the extra dimensions\footnote{Hence, this parallelism excludes distances, in the horizontal and gauge-invariant sense quantified by torsion. This might encourage to call the flat bundle ''absolutely parallel'', to make the distinction with the ''teleparallel'' - i.e. parallel at a distance - where flatness is not required in the homogeneous sector and nonzero displacement curvature field strengths are tolerated. We will not adopt that terminology here since symmetric teleparallel geometry is descriptive enough. We will refer to the gauge in symmetric teleparallel geometry, where the affine connection vanishes, as {\it the coincident gauge}, since it comes with even the inertial displacement field excluded. It will be shown in Section \ref{symsec} that the requirement of the coincident gauge to exist in any coordinate system singles out the Symmetric Teleparallel Equivalent of GR from the most general 5-parametric quadratic action. This statement will be made more precise later on, but the underlying reason for this is the existence of a second 4-parameter symmetry.}. 
We are led to investigate the symmetric teleparallel geometry, which was introduced by Nester and Yo in 1998  \cite{Nester:1998mp}. 
A flat connection needs not to be metric compatible, and then the dynamics of GR can be reproduced in a teleparallel geometry even without contortion, i.e. even in symmetric teleparallel geometry (STG). We will show that STG becomes simple to understand and work with in the Palatini formalism.  Though STG has been rather seldom studied \cite{Poltorak:2004tz,Mol:2014ooa},  
the embedding of the symmetric teleparallel theory into the metric-affine gauge
theory \cite{Hehl:1994ue} has been developed by Adak {\it et al}  \cite{Adak:2006rx}, who have also considered the coupling of spinor fields in STG and derived several exact solutions
to a general quadratic theory \cite{Adak:2004uh,Adak:2005cd,Adak:2008gd}. Lagrange multipliers are employed to impose the vanishing of torsion and curvature,
and we arrive, in the Palatini formalism, at a system of equations of motion that can be shown to be the coordinate manifestation of the metric-affine gauge theory in STG. A subtle completion perhaps, which
we find crucial to the interpretation of the theory, is the conservation
of energy-momentum and its direct relation with the Bianchi identities for theories formulated in the STG.
Specifically, this means that the metric divergence of the metric field equations vanishes, separately (on-shell) on the right hand side where it is known
as the continuity of matter or the conservation of energy-momentum, and (off-shell) on the left hand side where it is sometimes known as the generalised Bianchi identity \cite{Koivisto:2005yk}. To ensure this in a general theory, one has to take into account the inertial connection. Unless the action has an extra symmetry such that the inertial connection vanishes identically, which occurs for the  equivalent of GR in STG, one cannot in general trivialise both the connection and the metric simultaneously. This is completely analogous to the situation in the metric compatible context of torsion teleparallelism, and also here we expect that when the form of the action deviates from the equivalent of GR, some modes are promoted from gauge to physical. This article explores the general quadratic theory as an example and brings the symmetric teleparallel inertial connection to discussion \cite{BeltranJimenez:2017tkd,Conroy:2017yln}. 
 
We begin by reviewing the Palatini formalism in section \ref{palatini}. The notation is introduced for the basic geometrical objects and we formally derive the field equations before yet specifying an action.
In section \ref{telesec} we restrict the theory to the metric compatible and flat geometry. As the action, we consider a general parity-even theory that is quadratic in derivatives and in torsion.
The less explored STG is considered in section \ref{symsec}. Again, for concreteness we demonstrate workings of the formalism with a general parity-even 
theory that is now quadratic in non-metricity instead. As an example, the equations are adapted to the isotropic and homogeneous cosmological background. The formalism is applied to the context of bootstrap of the perturbative field theory, and to the considerations of the Euclidean action that can be regular in STG. We only point out these as interesting applications, without aiming to finalise them, 
in section \ref{sec:applications}. We summarise the results and consider directions for further studies in section \ref{concsec}. 
 
\section{Palatini theories}
\label{palatini}

To set up the notations and conventions, we will start by briefly reviewing the basic geometrical objects that will be used throughout this work. From the geometric objects defined in \ref{psub1}, we will then construct an invariant action and derive field equations in \ref{psub2}, by varying wrt the two fundamental fields in the Palatini formalism: the metric and the connection.

\subsection{Geometrical framework}
\label{psub1}

Let us first introduce the affine connection $\Gamma^\alpha{}_{\mu\nu}$ that determines the covariant derivative of tensors. It can be defined by its action on vectors $V^\alpha$ and co-vectors $V_\alpha$,
\ba \label{covariants}
\nabla_\mu V^\alpha & = & V^\alpha_{\phantom{\alpha},\mu} + \Gamma^\alpha_{\phantom{\alpha}\mu\lambda}V^\lambda\,, \\
\nabla_\mu V_\alpha & = & V_{{\alpha},\mu} - \Gamma^\lambda_{\phantom{\alpha}\mu\alpha}V_\lambda\,.
\ea
These expressions also determine the action of the covariant derivative on arbitrary tensors, which follows by the direct generalisation of the above. Given a connection, the commutator of covariant derivatives defines important objects related to the intrinsic properties of the space. In the present case, we can compute such a commutator action on a scalar $\phi$ to obtain 
\be\label{DDT}
\big[\nabla_\mu, \nabla_\nu\big]\phi=-T^\lambda{}_{\mu\nu}\partial_\lambda\phi\,,
\ee
where we have defined the torsion tensor as
\be \label{torsion}
{T}^\alpha_{\phantom{\alpha}\mu\nu} = 2\Gamma^\alpha_{\phantom{\alpha}[\mu\nu]}\,,
\ee
i.e., it is identified with the antisymmetric piece of the connection. Analogously, the action of the commutator upon a vector field is
\be \label{geometry}
\big[\nabla_\mu,\nabla_\nu\big]V^\alpha = {R}^\alpha_{\phantom{\alpha}\beta\mu\nu}V^\beta - {T}^\beta_{\phantom{\beta}\mu\nu}\nabla_\beta V^\alpha\,,
\ee
where the curvature Riemann tensor is defined as
\be \label{riemann}
{R}^\alpha_{\phantom{\alpha}\beta\mu\nu} = 
2\partial_{[\mu} \Gamma^\alpha_{\phantom{\alpha}\nu]\beta}
+ 2\Gamma^\alpha_{\phantom{\alpha}[\mu\lvert\lambda\rvert}\Gamma^\lambda_{\phantom{\lambda}\nu]\beta}\,.
\ee
Applying the commutator to higher rank tensors does not generate new objects so we already have the two fundamental tensors associated with the connection. We can now introduce the following trace of the Riemann tensor\footnote{For a general connection one can introduce the independent trace $R^\alpha{}_{\alpha\mu\nu}$ sometimes known as homothetic tensor and, in the presence of a metric, one can introduce yet another independent trace, the co-Ricci tensor $g^{\alpha\beta}R^\mu{}_{\alpha\beta\nu}$. We will not make use of these additional objects in this work so their existence will not be relevant for us, but let us point out that the trace of the co-Ricci tensor coincides with the Ricci scalar up to a sign, and the homothetic tensor describes the non-metric curvature.} 
\be \label{ricci}
{R}_{\mu\nu} ={R}^\alpha_{\phantom{\alpha}\mu\alpha\nu}\,,
\ee
which defines the Ricci tensor in the usual manner.

In addition to the affine structure defined by the connection, we can incorporate a metric structure, defined by a metric tensor with the components $g_{\mu\nu}$. Then, we can construct the Ricci scalar defined as
\be
R = g^{\mu\nu}{R}_{\mu\nu}\,. 
\ee
Once the metric structure $g_{\mu\nu}$ is introduced besides the affine connection $\Gamma$, the metric singles out a special connection given by the unique torsion-free connection compatible with it, in the sense that the metric is covariantly constant. This connection goes under the name of Levi-Civita connection and is given by the Christoffel symbols
\be \label{christoffel}
\left\{^{\phantom{i} \alpha}_{\beta\gamma}\right\} = \frac{1}{2}g^{\alpha\lambda}\lp g_{\beta\lambda,\gamma}
+  g_{\lambda\gamma,\beta} - g_{\beta\gamma,\lambda}\rp\,.
\ee
We will denote the covariant derivative wrt this connection by $\mathcal{D}_\mu$, so we have $\mathcal{D}_\alpha g_{\mu\nu}=0$ by definition.
When torsion is allowed, the full connection can be written as the above Levi-Civita part plus the contortion piece due to (\ref{torsion}),
\be \label{contortion}
K^\alpha_{\phantom{\alpha}\mu\nu} = \frac{1}{2}T^\alpha_{\phantom{\alpha}\mu\nu} + T_{(\mu{\phantom{\alpha}\nu)}}^{\phantom{,\mu}\alpha}\,.
\ee 
The contortion is notorious for its antisymmetry in the first and last indices, which implies the vanishing of the trace $K^\alpha_{\phantom{\alpha}\mu\alpha}=0$, while its other traces are given in terms of the torsion trace vector $T_\mu=T^\alpha_{\phantom{\alpha}\mu\alpha}$ as 
\be
K^\alpha_{\phantom{\alpha}\alpha\mu}=-K_{\mu\phantom{\alpha}\alpha}^{\phantom{\mu}\alpha}=-T_\mu\,.
\ee
One related object that will be relevant for our study later is what we will refer to as the conjugate torsion, defined by
\be
{S}{}_\alpha^{\phantom{\alpha}\mu\nu}   =  a T_\alpha^{\phantom{\alpha}\mu\nu} + b T^{[\mu\phantom{\alpha}\nu]}_{\phantom{,\mu}\alpha} + c\delta^{[\mu}_\alpha T^{\nu]}\,, \label{super}
\ee
where $a$, $b$ and $c$ can be arbitrary constants. This object can be considered as a generalisation of the superpotential in teleparallel gravity models\footnote{Which can be obtained
as a generalisation of the usual Hodge dual of the torsion two-form \cite{Lucas:2008gs,Huang:2014gta}.}, defined as
\be
\mathring{S}_\alpha^{\phantom{\alpha}\mu\nu} =  K^{\mu\phantom{\alpha}\nu}_{\phantom{\mu}\alpha} + \delta^{\mu}_\alpha T^{\nu} - \delta^{\nu}_\alpha T^{\mu}\,, \label{supergr}
\ee
to the three-parameter form that takes into account the independent even contractions with the torsion tensor \cite{PELLEGRINI:19165,Hayashi:1979qx,Mielke:1992te}, and reduces to a one-parameter generalisation in the case of the New GR\footnote{The TEGR is also sometimes referred to New GR.} \cite{Hayashi:1979qx}. 
The expression (\ref{supergr}) is the term appearing in TEGR \cite{Aldrovandi:2013wha,Maluf:2013gaa}, but in all our formulae we will consider the general conjugate $S_\alpha{}^{\mu\nu}$ given in (\ref{super}) and, in particular, we will define the torsion scalar as
\ba \label{mathbbt}
\mathbb{T} &= & \frac{1}{2}S_\alpha^{\phantom{\alpha}\mu\nu}T^\alpha_{\phantom{\alpha}\mu\nu} \nn \\ & = &  \frac{1}{2}\lp aT_{\alpha\mu\nu}+bT_{\mu\alpha\nu}+cg_{\alpha\mu}T_\nu\rp T^{\alpha\mu\nu}\,,
\ea 
unless otherwise stated. The corresponding object in the limit of TEGR, given by $a=1/4$, $b=1/2$ and $c=-1$, we then denote as
\be
\mathring{\mathbb{T}} = \frac{1}{2}\mathring{S}_\alpha^{\phantom{\alpha}\mu\nu}T^\alpha_{\phantom{\alpha}\mu\nu}.
 \ee
The curvature and the torsion tensors components are not completely independent since they satisfy a set of relations derived from the Jacobi identity for the covariant derivative, 
\be
[\nabla_\alpha,[\nabla_\beta,\nabla_\gamma]] + [\nabla_\beta,[\nabla_\gamma,\nabla_\alpha]] + [\nabla_\gamma,[\nabla_\alpha,\nabla_\beta]]=0\,,
\ee
which results in the Bianchi identities  
\ba
R^\alpha_{\phantom{\alpha}\beta(\mu\nu)}&= & 0\,,  \label{Bianchi1} \\
R^\mu_{\phantom{\mu}[\alpha\beta\gamma]} - \nabla_{[\alpha}T^\mu_{\phantom{\mu}\beta\gamma]} + T^\nu_{\phantom{\nu}[\alpha\beta}T^\mu_{\phantom{\mu}\gamma]\nu} & = & 0\,, \label{Bianchi2}\\
\nabla_{[\alpha}R^{\mu}_{\phantom{\mu}\lvert\nu\rvert\beta\gamma]} - T^\lambda_{\phantom{\lambda}[\alpha\beta}R^{\mu}_{\phantom{\mu}\vert\nu\rvert\gamma]\lambda} & = & 0  \label{Bianchi3}\,.
\ea
The first identity is a direct consequence of the definition of the Riemann tensor by (\ref{riemann}), while the second and the third identities are a consequence of applying the Jacobi identity to a scalar and a vector respectively.

An additional important tensor that has not appeared yet is the so called non-metricity of the connection defined by 
\be \label{nonm}
Q_{\alpha\mu\nu} = \nabla_\alpha g_{\mu\nu}\,,
\ee 
that gives the failure of the connection in being metric compatible. We can alternatively write the above relation in terms of the metric inverse as 
\be
Q_\alpha{}^{\mu\nu}=-\nabla_\alpha g^{\mu\nu}.
\ee
The non-metricity contributes to the connection the {\it  disformation} term given by 
\be \label{disformation}
L^\alpha_{\phantom{\alpha}\mu\nu}  = \frac{1}{2} Q^{\alpha}_{\phantom{\alpha}\mu\nu} - Q_{(\mu\phantom{\alpha}\nu)}^{\phantom{(\mu}\alpha}\,. 
\ee
The full affine connection can thus be split in the following three pieces 
\be \label{decomposition}
\Gamma^\alpha_{\phantom{\alpha}\mu\nu}=\left\{^{\phantom{i} \alpha}_{\mu\nu}\right\} + K^\alpha_{\phantom{\alpha}\mu\nu} + L^\alpha_{\phantom{\alpha}\mu\nu}\,,
\ee
that correspond to the Levi-Civita part, the contortion part and the disformation generated by the non-metricity. The non-metricity satisfies
\be\label{Qbianchi}
\nabla_{[\mu}Q_{\nu]\alpha\beta} = R_{(\alpha\beta)\nu\mu}-\frac12T^\lambda{}_{\mu\nu} Q_{\lambda\alpha\beta} \,,
\ee
as one can verify by direct computation, for instance using the general expression for the commutator of two covariant derivatives acting on the metric tensor. This relation shows how the Riemann tensor becomes antisymmetric also in the first pair of indices in spacetimes with vanishing non-metricity.
Also by direct calculation using the definition (\ref{riemann}) one can verify that, if the connection is shifted by an arbitrary tensor as
$\hat{\Gamma}^\alpha_{\phantom{\alpha}\mu\nu} = {\Gamma}^\alpha_{\phantom{\alpha}\mu\nu} + \Omega^\alpha_{\phantom{\alpha}\mu\nu}$, 
then the Riemann tensor changes as
\be \label{hatR}
 \hat{R}^\alpha_{\phantom{\alpha}\beta\mu\nu} =R^\alpha_{\phantom{\alpha}\beta\mu\nu} + T^\lambda_{\phantom{\lambda}\mu\nu}\Omega^\alpha_{\phantom{\alpha}\lambda\beta} + 2\nabla_{[\mu}\Omega^\alpha_{\phantom{\alpha}\nu]\beta} +2\Omega^\alpha_{\phantom{\alpha}[\mu\lvert\lambda\rvert}\Omega^\lambda_{\phantom{\lambda}\nu]\beta}\,.
\ee
This expression can then be used to relate the curvatures of the Levi-Civita connection and that of a general connection featuring torsion and/or non-metricity. 

We will end this brief discussion of geometrical objects by giving some relations between the Levi-Civita connection and the general one. We will denote the components of the Riemann and Ricci curvature of the Levi-Civita connection (\ref{christoffel}) as $\mathcal{R}^\alpha_{\phantom{\alpha}\beta\mu\nu}$ and $\mathcal{R}_{\mu\nu}$, respectively. The metric Ricci scalar is then $\mathcal{R}=g^{\mu\nu}\mathcal{R}_{\mu\nu}$. Then, applying the formula (\ref{hatR})
in the case of metric-compatible connection with $Q_{\alpha\mu\nu}=0$, we obtain that
\be
R_{\mu\nu} = \mathcal{R}_{\mu\nu} + \mathcal{D}_\alpha\lp \mathring{S}^{\phantom{\nu}\alpha}_{\nu\phantom{\alpha}\mu} + g_{\mu\nu} T^\alpha\rp - T_\alpha K^\alpha_{\phantom{\alpha}\nu\mu} - K_{\alpha\nu\beta}K^{\beta\alpha}_{\phantom{\beta\alpha}\mu}\,,
\ee
where we have used the covariant Levi-Civita derivative $\mathcal{D}_\mu$, and recall the superpotential $\mathring{S}^\alpha_{\phantom{\alpha}\nu\mu} $ specified in (\ref{supergr}). Contracting the indices gives the celebrated relation of the curvatures,
\be\label{relationRT}
R = \mathcal{R} + \mathring{\mathbb{T}} + 2\mathcal{D}_\alpha T^\alpha\,.
\ee 
One can then see that the Einstein field equation of GR (with a source $\mathfrak{T}_{\mu\nu}$ to be defined later)
can be written, in the teleparallel gauge where $R^\alpha_{\phantom{\alpha}\beta\mu\nu}=0$, solely in terms
of the torsion instead of curvature
\ba
\mathfrak{T}_{\mu\nu} & = &  \mathcal{R}_{\mu\nu} - \frac{1}{2}g_{\mu\nu} \mathcal{R} \quad =  \nonumber \\
  \mathcal{D}_\alpha \mathring{S}_{\nu\mu}^{\phantom{\nu\mu}\alpha} & + & \lp T_{\alpha\beta\mu}-K_{\alpha\beta\mu}\rp \mathring{S}^{\alpha\beta}_{\phantom{\alpha\beta}\nu} + \frac{1}{2}g_{\mu\nu}\mathring{\mathbb{T}}
 \,.
\ea
In the next section \ref{telesec} we will derive this equation from the teleparallel Palatini action and verify that it can be generalised to arbitrary quadratic theories straightforwardly by setting $\mathring{\mathbb{T}} \rightarrow \mathbb{T}$ and $\mathring{S}_{\nu\mu}^{\phantom{\nu\mu}\alpha} \rightarrow {S}_{\nu\mu}^{\phantom{\nu\mu}\alpha}$ in the above.

\subsection{Field equations}
\label{psub2}

For the sake of generality, here we will derive the field equations for an arbitrary action formulated in a (holonomic) metric-affine framework where the connection and the metric are regarded as independent entities, which usually receives the name of Palatini formalism (see footnote \ref{footnote}). Let us consider a theory described by the following gravitational Lagrangian 
\be
\mathcal{L}_G  =\sqrt{-g}f(g^{\mu\nu},R^\alpha_{\phantom{\alpha}\beta\mu\nu},T^\alpha_{\phantom{\alpha}\mu\nu})\,,
\ee
where $f$ is a scalar function constructed by forming invariants with the curvature and the torsion, using the metric tensor for the contractions\footnote{Most generally, one could also take into account
 the connection (\ref{christoffel}) generated by the metric. This would lead to the hybrid metric-Palatini gravity \cite{Capozziello:2015lza,Leanizbarrutia:2017xyd,VargasdosSantos:2017ggl}. Hybrid teleparallel gravity would be a possible extension of the present framework. \label{hybrid}}. 
In addition to the gravitational Lagrangian we will take into account a matter sector that will act as a source of the gravitational equations, i.e., its Lagrangian will be $\mathcal{L}_M[\Psi,g_{\mu\nu},{\Gamma}]$ with a coupling of the matter fields $\Psi$ to the metric and the connection. Therefore, the class of  theories that we consider will be described by the Lagrangian $\mathcal{L}=\mathcal{L}_G+\mathcal{L}_M$. The contribution of this matter sector to the gravitational equations will be by means of the energy-momentum tensor $\T_{\mu\nu}$ and the hypermomentum tensor density $\mathfrak{H}_\alpha{}^{\mu\nu}$ defined as
\be 
\T_{\mu\nu} = -\frac{2}{\sqrt{-g}}\frac{\delta \mathcal{S}_M }{\delta g^{\mu\nu}}\,, \quad 
\mathfrak{H}_\lambda{}^{\mu\nu} = -\frac12\frac{\delta \mathcal{S}_M}{\delta \Gamma^\lambda_{\phantom{\lambda}\mu\nu}}\,,
\ee
where $\mS_M=\int\diff^4x\mL_M$ is the corresponding matter action. Notice that, while the energy-momentum tensor is a tensor, the hypermomentum is defined as a tensorial density of weight\footnote{Let us remember that a tensorial density of weight $w$ is an object that transforms like a tensor except that it picks an extra power $w$ of the Jacobian. Thus, for instance $\sqrt{-g}$ is a tensorial scalar with $w=-1$.} $-1$.

The metric field equations can be easily computed and are given by 
\be \label{einsteintensors}
2\frac{\partial f}{\partial g^{\mu\nu}} - f g_{\mu\nu} =  \T_{\mu\nu}\,.
\ee
The field equations for the connection require a little more effort and some care when integrating by parts. To alleviate the notation, it is helpful to introduce the following two conjugate densities
 of the function $f$ wrt the curvature and the torsion:
\ba 
f_\alpha^{\phantom{\alpha}\beta\mu\nu} & = & \sqrt{-g}\frac{\partial f}{\partial  R^\alpha_{\phantom{\alpha}\beta\mu\nu}}\,, \label{density1} \\
f_\alpha^{\phantom{\alpha}\mu\nu} & = & \sqrt{-g}\frac{\partial f}{\partial {T}^\alpha_{\phantom{\alpha}\mu\nu}} \label{density2}\, .
\ea
Notice that these objects will have the antisymmetry in their last two indices inherited from the curvature and the torsion. Similarly to the hypermomentum, these conjugate quantities have been defined as tensorial densities of weight $-1$. When integrating by parts one has to take into account that the connection can be incompatible with the metric as well as the presence of torsion. Moreover, another subtlety is that such integrations by parts will be performed for vectorial densities and we need to recall that, for a vectorial density $f^\mu$ of arbitrary weight $w$ we have 
\begin{align}
\nabla_\mu f^\mu=&\partial_\mu f^\mu+\Gamma^\mu_{\mu\alpha}f^\alpha+w\Gamma^\mu_{\alpha\mu}f^\alpha\\
=&\partial_{\mu} f^\mu - \left[T_\mu + \lp 1+w\rp\lp -\left\{^{\phantom{i} \alpha}_{\mu\alpha}\right\} +\frac12Q_{\mu}\rp\right]f^\mu\,,\nonumber\\
=& \partial_{\mu} f^\mu - \left[T_\mu + \frac{1}{2}\lp 1+w\rp\lp g^{\alpha\beta}g_{\alpha\beta,\mu}+Q_{\mu}\rp\right]f^\mu\,,\nonumber
\end{align}
with $Q_\mu=Q_{\mu\phantom{\alpha}\alpha}^{\phantom{\mu}\alpha}$ so the non-metricity part vanishes for $w=-1$ even when $Q_{\alpha\mu\nu} \neq 0$. This is indeed the case one finds when varying the action so that integration by parts will generate extra contributions proportional to the torsion.

After introducing the appropriate notation and clarifying some subtleties we are now ready to obtain the connection field equations. For that we will use the variation of the Riemann tensor easily deduced from the infinitesimal version of (\ref{hatR})
\be
\delta_\Gamma R^\alpha_{\phantom{\alpha}\beta\mu\nu} = 
2\nabla_{[\mu}  \delta\Gamma^{\alpha}_{\phantom{\alpha}\nu]\beta} 
+ T^\gamma_{\phantom{\gamma}\mu\nu}\delta\Gamma^{\alpha}_{\phantom{\alpha}\gamma\beta}\,,
\ee
and the variation of the torsion 
\be
\delta_\Gamma {T}^\alpha_{\phantom{\alpha}\mu\nu} = 2\delta \Gamma^\alpha_{\phantom{\alpha}[\mu\nu]}.
\ee 
The total variation of the gravitational Lagrangian then reads 
\be \label{gammavar}
\delta_\Gamma \mathcal{L}_G  = f_\alpha^{\phantom{\alpha}\beta\mu\nu}  \delta_\Gamma R^\alpha_{\phantom{\alpha}\beta\mu\nu} 
+  f_\alpha^{\phantom{\alpha}\mu\nu}\delta_\Gamma {T}^\alpha_{\phantom{\alpha}\mu\nu}\,,
\ee
which, upon use of the above formulae, leads to the connection field equations 
\be \label{eomriemann}
\Big(\nabla_\beta+T_\beta\Big) f_{\alpha}^{\phantom{\alpha}\nu\mu\beta} + 
\text{\footnotesize $\frac{1}{2}$}T^\mu_{\phantom{\nu}\beta\gamma}f_{\alpha}^{\phantom{\alpha}\nu\beta\gamma}
 + f_\alpha^{\phantom{\alpha}\mu\nu}
= \mathfrak{H}_\alpha^{\phantom{\alpha}\mu\nu}\,.
\ee
If $2f=g^{\mu\nu}R_{\mu\nu}$ and we neglect torsion, this reduces to $\nabla_\alpha \lp \sqrt{-g} g^{\mu\nu} \rp = 2\mathfrak{H}_\alpha^{\phantom{\alpha}\mu\nu}$.
In the absence of hypermomentum we recover the metric-compatible connection (\ref{christoffel}) and the dynamics of GR. When we take the torsion into account, 
after some manipulations we still find that the connection is metric-compatible, but it contains a projective mode (see below) of contortion. Since this vectorial mode decouples from the
dynamics they nevertheless remain equivalent with GR.  

In the Palatini formalism without the teleparallel restriction, the torsion is typically an auxiliary field. 
This is manifested for example in the Einstein-Cartan-Kibble-Sciama theory \cite{Hehl:1976kj} and its supergravitating extensions,
where the presence of torsion leads to the well-known four-fermion contact interaction, the Riemann curvature being the gravitational field strength as usual \cite{Hehl:1976kj}.   
However, in the teleparallel Palatini formalism, the dynamics of GR can be reproduced by choosing a specific quadratic
combination of the torsion terms as we will see in the next section. The stress energy and the hypermomentum of matter, the latter
being generated by coupling fermions to gravitation, become unified into a source term appearing in the generalised field equation (\ref{efes3}) which will be a main
result of the first part of this article. The hypermomentum will then enter via a derivative term.

Now that we have introduced the geometrical framework and derived the field equations for the general case we can proceed with the particular cases of interest for our study.

\section{Metric Teleparallelism}
\label{telesec}

Let us consider the quadratic torsion action given by the scalar defined in (\ref{mathbbt}) with the teleparallelity and metricity constraints imposed by appropriate Lagrange multipliers:
\be \label{abcd}
\mathcal{L}_G   =  \frac12\sqrt{-g}\mathbb{T}  
+ 
\lambda_\alpha^{\phantom{\alpha}\beta\mu\nu} R^\alpha_{\phantom{\alpha}\beta\mu\nu} + \lambda^\alpha_{\phantom{\alpha}\mu\nu}\nabla_\alpha g^{\mu\nu}\,.
\ee
We have two Lagrange multipliers, a rank-4 tensor density with the symmetry $\lambda_{\alpha}^{\phantom{\alpha}\mu\beta\nu}  = \lambda_{\alpha}^{\phantom{\alpha}\mu[\beta\nu]}$,
and a rank-3 tensor density with the symmetry $\lambda^\alpha_{\phantom{\alpha}\mu\nu}=\lambda^\alpha_{\phantom{\alpha}(\mu\nu)}$, both of them with weight $-1$. The curvature conjugate (\ref{density1}) is directly given by the Lagrange multiplier $\lambda_{\alpha}^{\phantom{\alpha}\mu\beta\nu}$. On the other hand,
recalling the definition of $S_\alpha^{\phantom{\alpha}\mu\nu}$ in (\ref{super}), the conjugate torsion (\ref{density2}) for the Lagrangian (\ref{abcd}) reads
\be \label{density3}
{f_{\alpha}^{\phantom{\alpha}\mu\nu}} = \frac12 {\sqrt{-g}} S_\alpha^{\phantom{\alpha}\mu\nu}\,,
\ee
with its trace, in an arbitrary dimension $n$, given by
\be \label{tracetorsiontrace}
f_\alpha{}^{\mu\alpha}= \frac{1}{4}{\sqrt{-g}}\big[ 2a + b - (n-1) c \big] T^\mu\,.
\ee
From this expression we see that the trace vanishes for theories with $2a+b-(n-1)c=0$. In that case, the torsion scalar $\mathbb{T}$ enjoys a projective symmetry $\Gamma^\alpha_{\phantom{\alpha}\mu\beta}\rightarrow \Gamma^\alpha_{\phantom{\alpha}\mu\beta}+A_\mu(x)\delta^\alpha{}_\beta$ with $A_\mu(x)$ an arbitrary $1$-form field. Then the vanishing of the above trace can be understood as a consequence of this symmetry. In general metric-affine theories, the presence of this symmetry allows to remove the vector trace of the torsion, which only enters as a projective mode. In the theory described by (\ref{abcd}), this projective symmetry is broken by the terms with the Lagrange multipliers\footnote{The breaking of the projective symmetry by suitable Lagrange multipliers was argued in \cite{Hehl:1981ed} to be necessary to solve a consistency problem in the Palatini-Hilbert formulation of GR (which is projectively invariant) when the matter sector has hypermomentum. However, all standard matter fields, both bosonic and fermionic, respect the projective symmetry so no consistency problem arises. Moreover, even a matter field sector without such a symmetry does not necessarily leads to inconsistencies, but to some constraint equations which can be perfectly consistent, even desirable in some cases. }.  

The metric field equations (\ref{einsteintensors}), for the models (\ref{abcd}) are in this case
\be
\T_{\mu\nu}=\frac{\partial \mathbb{T}}{\partial g^{\mu\nu}}-\frac12\mathbb{T} g_{\mu\nu}-\frac{2}{\sqrt{-g}}\lp \nabla_\alpha + T_{\alpha}\rp \lambda^{\alpha}_{\phantom{\alpha}\mu\nu}\,,
\ee
whose form can be spelled out as 
\ba \label{fieldeq}
\T_{\mu\nu} & = & 
\frac{a}{2}\lp 2T_{\alpha\mu\beta} T^{\alpha\phantom{\nu}\beta}_{\phantom{\alpha}\nu} 
- T_{\mu\alpha\beta} T_\nu^{\phantom{\mu}\alpha\beta}- \frac{1}{2}T_{\alpha\beta\gamma}T^{\alpha\beta\gamma}g_{\mu\nu}\rp \nonumber \\
& + & \frac{b}{2}\lp T_{\alpha\mu\beta}T^{\beta\phantom{\nu}\alpha}_{\phantom{\alpha}\nu} - \frac{1}{2}T_{\alpha\beta\gamma}T^{\beta\alpha\gamma} g_{\mu\nu}\rp\nonumber \\
 & - & \frac{c}{2}\lp  T_\mu T_\nu - \frac{1}{2}T_\alpha T^\alpha g_{\mu\nu}\rp  \nonumber \\
 & - & \frac{2}{\sqrt{-g}}\lp \nabla_\alpha + T_{\alpha}\rp \lambda^{\alpha}_{\phantom{\alpha}\mu\nu}\,. 
\ea
It is useful to notice that this can be rewritten in the more compact form
\ba
&&T_{\alpha\beta\mu}S^{\alpha\beta}_{\phantom{\alpha\beta}\nu}-\frac12T_{\nu\alpha\beta}S_{\mu}^{\phantom{\mu}\alpha\beta} -\frac12g_{\mu\nu}\mathbb{T} \nn\\ &&\quad=\mathfrak{T}_{\mu\nu} +  \frac{2}{\sqrt{-g}}\lp \nabla_\alpha + T_{\alpha}\rp \lambda^{\alpha}_{\phantom{\alpha}\mu\nu}\,. \label{efes}
\ea
The equation (\ref{fieldeq}) is symmetric as it should and so is (\ref{efes}), although in a slightly less obvious way.
We can see from these equations that only the Lagrange multiplier enters with derivatives. However, there are other sets of equations to take into account. It will be shown in the following that the 
connection equation (\ref{eomriemann}) can be used to determine this Lagrange multiplier (or rather, its divergence appearing in (\ref{fieldeq})) so we can get rid of it in the metric field equations, being the effect of a source term involving the hypermomentum. 

The connection equations can be expressed as
\ba \label{eomlambda}
\Big(\nabla_\rho+T_\rho\Big) \lambda_{\alpha}{}^{\nu\mu\rho}+\frac12 T^\mu{}_{\rho\sigma}\lambda_\alpha{}^{\nu\rho\sigma}=\Delta_\alpha{}^{\mu\nu}\,,
\ea
where we have defined the source term
\be \label{delta}
\Delta_\alpha{}^{\mu\nu}\equiv \mathfrak{H}_\alpha{}^{\mu\nu}+f_\alpha{}^{\nu\mu}-\lambda^{\mu\nu}{}_\alpha\,,
\ee
where $f_{\alpha}^{\phantom{\alpha}\mu\nu}$ is given by (\ref{density3}) and we have used its anti-symmetry as well as the symmetry of $\lambda^\alpha{}_{\mu\nu}$. Besides the equations for the gravitational fields encoded in the metric and the connection, we have the constraint equations imposed by the Lagrange multiplier fields. The 4-index Lagrange multiplier 
restricts the dynamics of the connection by imposing teleparallelism:
\be \label{tele}
R^\alpha_{\phantom{\alpha}\beta\mu\nu} = 0\,.
\ee
To solve this equation, we note that  the affine curvature (\ref{riemann}) is the gauge field strength of the connection for general linear transformations $GL(4,\mathbb{R})$. We can use this fact to argue that, since Eq. (\ref{tele}) is trivially
solved by vanishing connection, then it must further be satisfied by any connection obtained by a general linear transformation. We will parameterise such a transformation by a
matrix $\Lambda^\alpha_{\phantom{\alpha}\beta}$, with $n^2$ independent components; the constraint (\ref{tele}) alone does not yet impose antisymmetry, because in the Palatini
variation the affine connection is not a priori restricted to a Lorentz connection. The general teleparallel Palatini connection, for which (\ref{tele}) must hold, is thus 
\be \label{lambda}
{\Gamma}^\alpha_{\phantom{\alpha}\mu\beta} = (\Lambda^{-1})^\alpha_{\phantom{\alpha}\nu}\partial_\mu \Lambda^\nu_{\phantom{\nu}\beta}\,,
\ee
implying that the teleparallel torsion is constrained to have the form
\be \label{tptorsion}
{T}^\alpha_{\phantom{\alpha}\mu\nu} = 2(\Lambda^{-1})^\alpha_{\phantom{\alpha}\beta}\partial_{[\mu} \Lambda^\beta_{\phantom{\beta}\nu]}\,.
\ee
The last piece of information comes from the equations of $\lambda^\alpha_{\phantom{\alpha}\mu\nu}$, which impose the metricity constraint
\be \label{nonm}
\nabla_\alpha g^{\mu\nu} =0\,.
\ee
This conveniently allows to raise and lower indices in our tensor equations, regardless of whether inside or outside covariant derivatives, as in the usual case. 
The equations (\ref{lambda}) and (\ref{nonm}) together imply the restriction 
\be \label{metricity}
g^{\lambda(\mu}\partial_\alpha\Lambda^{\nu)}{}_\rho(\Lambda^{-1})^\rho{}_\lambda =\frac{1}{2}\partial_\alpha g^{\mu\nu}{}\,,
\ee
which relates the derivatives of the metric and of the inertial connection field. 

We will now show how to obtain the divergence of $\lambda^\alpha{}_{\mu\nu}$ appearing in the metric field equations (\ref{fieldeq}) from the connection field equations (\ref{eomlambda}). To do that, we take the corresponding divergence of (\ref{eomlambda}). Due to the antisymmetry of $\lambda_\alpha{}^{\nu\mu\rho}$ in the last two indices we have
\be
\nabla_\mu\nabla_\rho\lambda_\alpha{}^{\nu\mu\rho}=\frac12\big[\nabla_\mu,\nabla_\rho\big]\lambda_\alpha{}^{\nu\mu\rho}=-\frac12 T^\kappa{}_{\mu\rho}\nabla_\kappa\lambda_\alpha{}^{\nu\mu\rho}\,,
\ee
where we have used the curvature-free condition (\ref{tele}). With this result, the divergence of (\ref{eomlambda}) simplifies to
\be \label{eomlambdaJ2}
\frac12\nabla_\mu T^\mu{}_{\rho\sigma}\lambda_\alpha{}^{\nu\rho\sigma}-\nabla_\mu T_\rho\lambda_\alpha{}^{\nu\rho\mu}-T_\rho\nabla_\mu\lambda_\alpha{}^{\nu\rho\mu}=\nabla_\mu\Delta_\alpha{}^{\mu\nu}\,.
\ee
We can now use (\ref{eomlambda}) to replace the covariant derivative $\nabla_\mu\lambda_\alpha{}^{\nu\rho\mu}$ so that (\ref{eomlambdaJ2}) can be expressed as
\be
\left[\frac12\Big(\nabla_\mu+T_\mu\Big)T^\mu{}_{\rho\sigma}+\nabla_{[\rho}T_{\sigma]}\right]\lambda_\alpha{}^{\nu\rho\sigma}=\Big(\nabla_\mu+T_\mu\Big)\Delta_\alpha{}^{\mu\nu}.
\ee
The term multiplying the Lagrange multiplier on the LHS of this equation can be seen to vanish by virtue of the Bianchi identities (\ref{Bianchi2}) for a curvature-free connection, since it exactly corresponds to the trace of those identities over $\alpha$ and $\mu$. We then obtain the equation
\be 
\Big(\nabla_\mu+T_\mu\Big)\Delta_\alpha{}^{\mu\nu}=0.
\label{eqDivlambda}
\ee
It can be useful to split this equation into symmetric and antisymmetric parts as
\ba 
\Big(\nabla_\mu+T_\mu\Big)\left[\lambda^\mu{}_{\alpha\nu}-{\mathfrak{H}}^{\phantom{(\nu}\mu}_{(\alpha\phantom{\mu}\nu)}-\frac12\sqrt{-g}S_{(\alpha\nu)}{}^\mu\right]=0\,, \\
\Big(\nabla_\mu+T_\mu\Big)\left[{\mathfrak{H}}^{\phantom{[\nu}\mu}_{[\alpha\phantom{\mu}\nu]}+\frac12\sqrt{-g}S_{[\alpha\nu]}{}^\mu\right]=0\,,
\label{eqDivlambda2}
\ea
where we see that the Lagrange multiplier decouples from the antisymmetric piece (as expected) so that only the symmetric part of this constraint will enter the metric field equations, as it will be confirmed shortly. 

Before proceeding further, let us notice that the relation (\ref{eqDivlambda}) that we have just obtained is simply a consequence of a gauge symmetry existing in the term $\mL_R=\lambda_\alpha{}^{\beta\mu\nu}R^\alpha{}_{\beta\mu\nu}$. As we already recalled, the Riemann tensor is nothing but the field strength of the connection of general linear transformations and, as such, it transforms covariantly under $GL(4,\mathbb{R})$ so that the term $\mL_R$ is invariant under the combined transformations
\ba
R^\alpha{}_{\beta\mu\nu}\rightarrow U^\alpha{}_\rho (U^{-1} )^\sigma{}_\beta R^\rho{}_{\sigma\mu\nu}\,,\\
\lambda_\alpha{}^{\beta\mu\nu}\rightarrow (U^{-1} )^\lambda{}_\alpha U^\beta{}_\kappa  \lambda_\lambda{}^{\sigma\mu\nu}\,,
\ea
where $U^\alpha{}_\beta\in GL(4,\mathbb{R})$. Under an infinitesimal transformation $U^\alpha{}_\beta=\delta^\alpha{}_\beta+J^\alpha{}_\beta$ the connection changes as usual $\delta\Gamma^\alpha_{\mu\beta}=-\nabla_\mu J^\alpha{}_\beta$ so that we have
\be
\delta\mL_R=R^\alpha{}_{\beta\mu\nu}\delta \lambda_\alpha{}^{\beta\mu\nu}-\frac{\delta\mL_R}{\delta\Gamma^\alpha_{\mu\nu}}\nabla_\mu J^\alpha{}_\nu.
\ee
Since the Lagrange multiplier equation imposes $R^\alpha{}_{\beta\mu\nu}=0$, the first term in this expression will vanish on-shell and, thus, it will not contribute to the conserved current associated to the symmetry. Integrating the covariant derivative in the second term by parts, taking into account that there is torsion and neglecting a boundary term, we finally obtain 
\be
\delta\mL_R=J^\alpha{}_\nu\Big(\nabla_\mu+T_\mu\Big)  \frac{\delta\mL_R}{\delta\Gamma^\alpha_{\phantom{\alpha}\mu\nu}}\,,
\ee
which gives the Bianchi identity 
\be 
\Big(\nabla_\mu+T_\mu\Big)  \frac{\delta\mL_R}{\delta\Gamma^\alpha_{\phantom{\alpha}\mu\nu}}=0\,,
\ee
that leads to (\ref{eqDivlambda}). Notice that this Bianchi identity is a consequence of the Riemann tensor being the curvature of the $GL(4,\mathbb{R})$ connection.

We can then return to the field equation (\ref{efes}). Using (\ref{eqDivlambda}) with the definition (\ref{delta}), we get
\ba
 \mathfrak{T}_{\mu\nu} & +&  \frac{2}{\sqrt{-g}}(\nabla_\alpha + T_\alpha)\mathfrak{H}^{\phantom{\mu}\alpha}_{(\mu\phantom{\alpha}\nu)} =  -(\nabla_\alpha + T_\alpha){S}_{(\mu\nu)}{}^\alpha \nn \\
& + &     T_{\alpha\beta\mu}S^{\alpha\beta}_{\phantom{\alpha\beta}\nu}-\frac12T_{\nu\alpha\beta}S_{\mu}^{\phantom{\mu}\alpha\beta}
 -  \frac12g_{\mu\nu}\mathbb{T}\,. \label{efes2}
\ea
where we have used the symmetry of $\lambda^\alpha_{\phantom{\alpha}\mu\nu}$ in the last two indices.

It is interesting to note that the divergence of the hypermomentum enters as an extra source term in these equations, i.e., the hypermomentum will act as an additional effective energy-momentum tensor. This appearance deserves some comments here or rather its disappearance in some cases. As usual, this extra piece will be determined by the way in which we decide to couple matter fields within this framework. The prescription for such couplings is closely related to what we should consider as minimally or non-minimally coupled matter fields. For standard scalar fields, the usual minimal coupling prescription stating that we should replace $\partial_\mu\rightarrow\nabla_\mu$ is unambiguous\footnote{At least for field theories with only first order derivatives. Again, ambiguities will arise in more general theories like Galileon or Horndeski type of Lagrangians \cite{Nicolis:2008in,Horndeski:1974wa}. Similar ambiguities will also arise in generalized Proca theories with broken gauge symmetry \cite{Heisenberg:2014rta,Jimenez:2016isa,Allys:2015sht}.} and, in fact, the absence of any coupling to the connection in this case will give rise to an identically vanishing hypermomentum. For abelian gauge fields we already encounter some ambiguities due to the presence of torsion. Following the same minimal coupling procedure one finds that the field strength results in $F_{\mu\nu}=2\nabla_{[\mu} A_{\nu]}=2\partial_{[\mu} A_{\nu]}-T^\lambda{}_{\mu\nu}A_\lambda$ leading to an explicit coupling to the torsion. In that case, the hypermomentum is non-trivial and its divergence will not vanish either so that there will be a contribution to the effective energy-momentum tensor through its divergence. However, a more appropriate minimal coupling prescription in theories with torsion would be to use the probably more meaningful definition of the field strength as the exterior derivative of the vector potential so that $F_{\mu\nu}=2\partial_{[\mu} A_{\nu]}$ directly and, thus, no hypermomentum will be generated. This definition is also preferable because it straightforwardly respects the $U(1)$ gauge symmetry (although couplings to torsion could be added while maintaining a $U(1)$ gauge symmetry involving the torsion).  Finally, fermions do couple to the connection in any case. Whether they couple to the torsion or not is, to some extent, a matter of choice, i.e., if we decide to couple them to the Cartan connection or to the Ehresmann connection. The latter
 follows naturally in the coset construction, and then a similar argument that leads to (\ref{eqDivlambda}) will lead to $(\nabla_\alpha + T_\alpha)\mathfrak{H}^{\phantom{\mu}\alpha}_{\mu\phantom{\alpha}\nu} = 0$ by virtue of the gauge invariance in the matter sector. As we see, although the minimal coupling prescription can be subtle, the most standard cases lead to a vanishing contribution from the hypermomentum to the effective energy-momentum tensor. However, we will keep it for the sake of generality and explore some consequences elsewhere.

After this little digression concerning the hypermomentum, we can proceed to obtain the final field equations. 
The divergence of the generalised superpotential can be related, by using (\ref{decomposition}), to the divergence w.r.t. 
the metric compatible derivative $\mathcal{D}_\mu$ with the symbol (\ref{christoffel}), as follows:
\ba
\nabla_\alpha S_{\mu\nu}{}^\alpha=&&\mathcal{D}_\alpha S_{\mu\nu}{}^\alpha-T_\alpha S_{\mu\nu}{}^\alpha\nonumber\\
&&-K^\lambda{}_{\alpha\mu}S_{\lambda\nu}{}^\alpha-K^\lambda{}_{\alpha\nu}S_{\mu\lambda}{}^\alpha\,.
\ea
We can now rewrite (\ref{efes2}) in terms of the metric compatible covariant derivative. It is important to keep in mind the symmetry of $\lambda^\alpha{}_{\mu\nu}$, due to which only the symmetric part of the above expression will be required. We finally find
\ba \label{mfe}
 \mathfrak{T}_{\mu\nu} & +&  \frac{2}{\sqrt{-g}}(\nabla_\alpha + T_\alpha)\mathfrak{H}^{\phantom{\mu}\alpha}_{(\mu\phantom{\alpha}\nu)} =  -\mathcal{D}_\alpha {S}_{(\mu\nu)}{}^\alpha \nn \\
& - &   S^{\alpha\beta}{}_\nu\left(T_{\alpha\mu\beta}+K_{\alpha\beta\mu} \right)
 -  \frac12g_{\mu\nu}\mathbb{T}\,. \label{efes2}
\ea
As a cross-check, after switching off the matter sources, we can write 
\be
\mathcal{D}_\mu S_\alpha^{\phantom{\alpha}\mu\beta}  +  S^{\mu\beta\lambda}\lp T_{\mu\alpha\lambda} + K_{\mu\lambda\alpha}\rp  -  \frac{1}{2}\delta^\beta_\alpha\mathbb{T}
 =   0\,,\,
\ee
to recover the field equation precisely in the form reported in \cite{Golovnev:2017dox}. Eventually, we can compactly summarise the metric field equations (\ref{mfe}) and the constraint (\ref{eqDivlambda2}) by removing the symmetrisation in the above equation, i.e., 
\ba
 \mathfrak{T}_{\mu\nu} & +&  \frac{2}{\sqrt{-g}}(\nabla_\alpha + T_\alpha)\mathfrak{H}^{\phantom{\mu}\alpha}_{\mu\phantom{\alpha}\nu} =  -\mathcal{D}_\alpha {S}_{\mu\nu}{}^\alpha \nn \\
& - &   S^{\alpha\beta}{}_\nu\left(T_{\alpha\mu\beta}+K_{\alpha\beta\mu} \right)
 -  \frac12g_{\mu\nu}\mathbb{T}\,, \label{efes3}
\ea
so that the symmetric part gives the metric field equations (\ref{mfe}), while its antisymmetric part gives the equation (\ref{eqDivlambda2}). One can show that the geometric part of that equation
vanishes identically when $S^\alpha{}_{\mu\nu} =  \mathring{S}^\alpha{}_{\mu\nu}$, by using the Bianchi identity (\ref{Bianchi2}).  

For completeness, we consider yet the possible symmetries of the two Lagrange multiplier tensor densities in the action. The freedom to redefine the Lagrange multipliers implies
an extra gauge symmetry of the theory. In the frame formalism of teleparallelism, the so called $\lambda$-symmetry
was clarified in the Hamiltonian formulation \cite{Blagojevic:2000qs,Blagojevic:2000pi}, see yet \cite{Nester:2017wau} for a subtle remark, and \cite{Obukhov:2002tm} for the additional
symmetry in the metric-affine extension of the Poincar\'e extension. In the present Palatini formulation, the analogous symmetries are of a different nature. In particular, the
possible redefinitions which leave the action invariant up to boundary terms, depend upon the gauge field strengths. We consider the constraint part of the Lagrangian density 
\be
\mathcal{L}_\lambda = \lambda^\alpha{}_{\mu\nu}Q_\alpha{}^{\mu\nu} + \lambda_\alpha{}^{\beta\mu\nu}R^\alpha{}_{\beta\mu\nu}\,. \label{lambda1}
\ee
Taking into account the obvious symmetries of the Lagrange multipliers $\lambda^\alpha{}_{\mu\nu}$ and $\lambda_\alpha{}^{\beta\mu\nu}$ inherited from those of the non-metricity and the Riemann, we see that they have 40 and 96 independent components respectively in four dimensions. However, by the use of the Bianchi identity (\ref{Bianchi3}), it is not 
difficult to see that if we introduce a rank-5 tensor density $\kappa_\alpha{}^{\beta\mu\nu\rho}$, which is totally antisymmetric in its three last indices, then
\be
\lambda_\alpha{}^{\beta\mu\nu} \rightarrow \lambda_\alpha{}^{\beta\mu\nu} + \lp \nabla_\rho + T_\rho\rp \kappa_\alpha{}^{\beta\mu\nu\rho} + T^\mu{}_{\rho\lambda}\kappa_\alpha{}^{\beta\lambda\nu\rho}\,,
\ee
results in $\mathcal{L}_\lambda \rightarrow \mathcal{L}_\lambda + \partial_\rho ( \kappa_\alpha{}^{\beta\mu\nu\rho}R^\alpha{}_{\beta\mu\nu})$. The tensor density $\kappa_\alpha{}^{\beta\mu\nu\rho}$ can have 40
independent components. Yet, we can introduce a rank-4 tensor density $\kappa^{\alpha\beta}{}_{\mu\nu}$, which is antisymmetric in its two first indices and symmetric in its two last indices, and has thus 60 independent components. We may then change both Lagrange multipliers in conjunction,
\ba
\lambda^\alpha{}_{\mu\nu} & \rightarrow & \lambda^\alpha{}_{\mu\nu} + \lp\nabla_\rho+T_\rho\rp \kappa^{\lambda\alpha}{}_{\mu\nu} \nn \\
& - & \frac{1}{2}T^\alpha{}_{\lambda\beta} \kappa^{\lambda\beta}{}_{\mu\nu} - 2Q_\lambda{}^\beta{}_{(\mu} \kappa^{\lambda\alpha}{}_{\nu)\beta}\,, \label{lambda2a} \\ 
\lambda_{\alpha\beta}{}^{\mu\nu} & \rightarrow & \lambda_{\alpha\beta}{}^{\mu\nu} -  \kappa^{\mu\nu}{}_{\alpha\beta}{}\,, \label{lambda2b}
\ea
such that the dynamics remain unaffected, since then $\mathcal{L}_\lambda \rightarrow \mathcal{L}_\lambda + \partial_\rho(\kappa^{\rho\beta}{}_{\mu\nu}Q_\alpha{}^{\mu\nu})$. This can be seen by using the
Bianchi identity (\ref{Qbianchi}).

\section{Symmetric teleparallelism}
\label{symsec}

The Palatini method can also be applied in torsion-free teleparallel theory. There will emerge, in conjuction with (\ref{lambda}), instead of the (\ref{metricity}), the condition $\Lambda^{\alpha}_{\phantom{\alpha}\nu}\partial_{[\mu}(\Lambda^{-1})^\nu_{\phantom{\nu}\beta]} = 0$. 
Given the dichotomy between curvature and torsion, one may choose neither, but then non-metricity has to be allowed in order to have a non-trivial connection. In the geometry that is flat in both sectors, i.e., vanishing curvature and torsion, there is still an inertial affine connection, but it can be made to vanish in the {\it coincident} gauge. As will be seen below, the requirement of the latter to exist at the level of the action singles out the Symmetric Teleparallel Equivalent of GR from the general 5-parametric quadratic action.

\subsection{Generalities}
The symmetric version of teleparallelism was suggested some time ago \cite{Nester:1998mp}. It has been studied in the language of differential forms in the context of metric-affine
gauge theories of gravitation \cite{Hehl:1994ue}. The field equations for the equivalent of GR have been obtained by transforming the Einstein field equations into the symmetric teleparallel gauge \cite{Adak:2004uh}, and more general field equations have been derived for a general action that is quadratic in non-metricity \cite{Adak:2005cd,Adak:2008gd}. An equivalent of  GR in STG was formulated also with differential forms, associating the coframe instead of the metric as the gravitational field in \cite{Mol:2014ooa}. Here we shall explore
the recently suggested  Palatini formulation \cite{BeltranJimenez:2017tkd} more systematically and fill in some details.

The non-metricity tensor can be decomposed into the sum of four irreducible components \cite{Hehl:1994ue,Obukhov:1997zd} as
\be
Q_{\alpha\mu\nu} = \sum_{i=1}^4 \prescript{(i)}{}Q_{\alpha\mu\nu}\,.
\ee
First. it is convenient to define the traces
\be \label{qtrace}
Q^\mu = Q^{\mu\alpha}_{\phantom{\mu\alpha}\alpha}\,, \quad \bar{Q}_\mu=\bar{Q}^\alpha_{\phantom{\alpha}\mu\alpha}\,,
\ee
where the traceless part has been separated as
\be
\bar{Q}_{\alpha\mu\nu}  =  Q_{\alpha\mu\nu} -  \frac{1}{4}g_{\mu\nu}Q_{\alpha}\,.
\ee
The vector components of non-metricity are then
\ba
 \prescript{(1)}{}Q_{\alpha\mu\nu} & = & \frac{1}{4}g_{\mu\nu}Q_\alpha\,, \\
 \prescript{(2)}{}Q_{\alpha\mu\nu} & = & \frac{2}{9}\lp g_{\alpha\mu}\bar{Q}_\nu + g_{\alpha\nu}\bar{Q}_\mu-\frac{1}{2}g_{\mu\nu}\bar{Q}_\alpha\rp\,. 
\ea
The tensor piece can be simply considered as 
\be
\bar{\bar{Q}}_{\alpha\mu\nu}  =  Q_{\alpha\mu\nu} - \prescript{(1)}{}Q_{\alpha\mu\nu}  - \prescript{(2)}{}Q_{\alpha\mu\nu}\,.
\ee
Its symmetric and antisymmetric pieces are irreducible (and writing them explicitly in components would
not be very helpful),
\ba
\prescript{(3)}{}Q_{\alpha\mu\nu} & = & \frac{1}{2}\lp \bar{\bar{Q}}_{\alpha\mu\nu} - \bar{\bar{Q}}_{\nu\mu\alpha} \rp \,, \\
 \prescript{(4)}{}Q_{\alpha\mu\nu} & = &  \frac{1}{2}\lp \bar{\bar{Q}}_{\alpha\mu\nu} + \bar{\bar{Q}}_{\nu\mu\alpha} \rp \,.
\ea
The action can then be specified by the five coefficients $c_i$, as
\ba 
\mathcal{L}_Q  & = & \sqrt{-g}\lp \sum_{i=1}^4 c_i  \prescript{(i)}{}Q^2 + c_5  \prescript{(3)}{}Q_{\alpha\mu\nu}  \prescript{(4)}{}Q^{\mu\alpha\nu}  \rp  \nonumber \\
& + &
\lambda_\alpha^{\phantom{\alpha}\beta\mu\nu} R^\alpha_{\phantom{\alpha}\beta\mu\nu} + \lambda_\alpha^{\phantom{\alpha}\mu\nu}T^\alpha_{\phantom{\alpha}\mu\nu}\,.
\label{abcd2a}
\ea
However, for us it is more convenient to rewrite the first line in a slightly different basis as
\ba \label{abcd2}
\mathcal{L}_G  & = & \frac{1}{2}\sqrt{-g} Q^\alpha_{\phantom{\alpha}\mu\nu}( c_1 Q_\alpha^{\phantom{\alpha}\mu\nu} 
 +  c_2 Q^{\mu\phantom{\alpha}\nu}_{\phantom{\mu}\alpha} 
+ c_3 g^{\mu\nu}Q_\alpha    \\ 
& + & c_4\delta^\mu_\alpha\tilde{Q}^\nu  + c_5\delta^\mu_\alpha {Q}^\nu) + \lambda_\alpha^{\phantom{\alpha}\beta\mu\nu} R^\alpha_{\phantom{\alpha}\beta\mu\nu} + \lambda_\alpha^{\phantom{\alpha}\mu\nu}T^\alpha_{\phantom{\alpha}\mu\nu}\,,\,\,\, \nn
\ea
where $\tilde{Q}^\nu = Q_\alpha^{\phantom{\alpha}\alpha\nu}$. 
Recall now the decomposition of the connection (\ref{decomposition}) and of its curvature (\ref{hatR}). Separating the disformation from the rest and tracing out two indices, we have,
when torsion is set to zero,
\ba \label{riccitensorq}
R_{\mu\nu} & = & \mathcal{R}_{\mu\nu} - L^\alpha_{\phantom{\alpha}\beta\mu}L^\beta_{\phantom{\alpha}\alpha\nu}  - \frac{1}{2}Q_\alpha L^\alpha_{\phantom{\alpha}\mu\nu} \nn \\
& + &   \mathcal{D}_\alpha L^\alpha_{\phantom{\alpha}\mu\nu} +   \frac{1}{2}\mathcal{D}_\nu Q_\mu \,.
\ea
Then we obtain the curvature scalar 
\be \label{ricciscalarq}
R   =  \mathcal{R}  + \mathcal{Q} +   \mathcal{D}_\alpha ( Q^\alpha - \tilde{Q}^\alpha )\,,
\ee
where the $\mathcal{Q}$ scalar is given by
\be \label{q2}
\mathcal{Q} =   \frac{1}{4}Q_{\alpha\beta\gamma}Q^{\alpha\beta\gamma} -  \frac{1}{2}Q_{\alpha\beta\gamma}Q^{\beta\gamma\alpha} 
  -   \frac{1}{4}Q_\alpha Q^\alpha  
  + \frac{1}{2}Q_\alpha\tilde{Q}^\alpha\,.
\ee
Note that the boundary term assumes a particularly simple form. In a flat geometry where $R=0$ we see that the Ricci curvature of the Levi-Civita connection $\mathcal{R}$ differs from $-\mathcal{Q}$ by a total derivative and, as a consequence, we conclude that the parameter values corresponding to the Einstein-Hilbert Lagrangian $f= \frac{1}{2}\mathcal{R}$ in (\ref{abcd2}) are thus
\be \label{qgr}
c_1 = -c_3= -\frac{1}{4}\,, \quad 
c_2 = -c_5= \frac{1}{2}\,, \quad
c_4 = 0\,.
\ee
The equivalent of GR in STG \cite{Nester:1998mp,Adak:2005cd,Adak:2008gd,Mol:2014ooa} is thus reproduced by this choice of parameters. We obtain here a result analogous to the one obtained for the metric teleparallelisms discussed in Sec. \ref{telesec} that a specific quadratic theory can reproduce the dynamics of GR up to a boundary term. Let us emphasise however that the boundary terms in the metric and symmetric teleparallels equivalents of GR are not the same or, in other words, these two equivalent formulations of GR also differ by a non-trivial boundary term between them.

There is another useful formula for the scalar $\mathcal{Q}$, which we find by recalling that non-metricity contributes to the connection (\ref{disformation}), and
by defining the scale connection $\hat{L}^{\alpha}_{\phantom{\alpha}\mu\nu}$ due to the trace of the $Q$-field as 
\ba
L^\alpha_{\phantom{\alpha}\mu\nu}  & = &  \frac{1}{2} Q^{\alpha}_{\phantom{\alpha}\mu\nu} - Q_{(\mu\phantom{\alpha}\nu)}^{\phantom{(\mu}\alpha}\,, \label{disformation2a} \\
\hat{L}^{\alpha}_{\phantom{\alpha}\mu\nu} & = & \frac{1}{2}g_{\mu\nu}Q^\alpha-\delta^\alpha_{(\mu}Q_{\nu)}\,. \label{disformation2b}
\ea
We note that since
\be 
L^\alpha_{\phantom{\alpha}\mu\alpha} = -\frac{1}{2}Q_\mu\,, \quad \hat{L}^\alpha_{\phantom{\alpha}\mu\alpha} = -\frac{n}{2}Q_\mu\,,
\ee
the combination losing one trace in $n$ dimensions is $L^\alpha_{\phantom{\alpha}\mu\nu} - \frac{2}{n}\hat{L}^\alpha_{\phantom{\alpha}\mu\nu}$. 
The scale connection is associated to the conformal class of the metric, whilst the traceless combination can be seen to represent the ''purely disformal'' part of the connection\footnote{In continuum mechanics, the change in the metric properties is often called ''deformation''. The $L^\alpha_{\phantom{\alpha}\mu\nu}$ is also known as ''distortion'', a term which we however reserve for the
$L^\alpha_{\phantom{\alpha}\mu\nu}+K^\alpha_{\phantom{\alpha}\mu\nu}$ \cite{Jimenez:2015fva,Jimenez:2016opp}.}.
The $\mathcal{Q}$-scalar (\ref{q2}) defining the action equivalent to GR is however
\be \label{qdef}
\mathcal{Q} = \frac{1}{2}Q_{\alpha\beta\gamma} \lp L^{\alpha\beta\gamma} - \hat{L}^{\alpha\beta\gamma}\rp\,.
\ee
The latter formula provides the starting point for a self-dual formulation to be discussed elsewhere.

To obtain the field equations for the generic quadratic theory, it is convenient to introduce the conjugate of the non-metricity 
\be \label{nmsuper}
\sqrt{-g}P^\alpha{}_{\mu\nu}  \equiv \frac{\partial\mL_Q}{\partial Q_\alpha{}^{\mu\nu}}\,,
\ee
for which we obtain, from (\ref{abcd2}), 
\ba \label{nmsuperquad}
P^\alpha{}_{\mu\nu} &=& c_1 Q^\alpha{}_{\mu\nu}+c_2Q_{(\mu\phantom{\alpha}\nu)}^{\phantom{\mu}\alpha}+c_3 Q^\alpha g_{\mu\nu} \nonumber \\
& + & c_4\delta^\alpha_{(\mu}\tilde{Q}_{\nu)} + \frac{c_5}{2}\big(\tilde{Q}^\alpha g_{\mu\nu}+\delta^\alpha_{(\mu}Q_{\nu)}\big)\,.
\ea
It may be worth noticing that, though $\mathcal{Q}=-Q_{\alpha\mu\nu}P^{\alpha\mu\nu}$, we have the relation
\be
P^\alpha{}_{\mu\nu}=-L^\alpha{}_{\mu\nu}+\hL^\alpha{}_{\mu\nu}-\frac12\Big(\tilde{Q}^\alpha g_{\mu\nu} -\delta^\alpha_{(\mu}Q_{\nu)}\Big),
\ee
instead of the one naively inferred from (\ref{qdef}). 

The field equations obtained by taking variations wrt the connection then gives
\be \label{ccEoM}
\nabla_\rho\lambda_\alpha{}^{\nu\mu\rho}+\lambda_\alpha{}^{\mu\nu}= \sqrt{-g}P^{\mu\nu}{}_\alpha+\mathfrak{H}_\alpha{}^{\mu\nu}\,,
\ee
where we have used the constraint of a torsion-free connection imposed by the Lagrange multiplier $\lambda_\alpha{}^{\mu\nu}$. 

Since the commutator of the covariant derivatives (\ref{geometry}) is proportional to the torsion and the curvature of the connection, we are free to commute
them in the STG. Consider first the covariant divergence of the previous equation\footnote{The role of equation (\ref{ccEoM}) is to determine
the curvature Lagrange multiplier and that of equation (\ref{cccEoM}) is to determine the torsion Lagrange multiplier. One could check the consistency of these 
equations by counting the degrees of freedom along the lines of \cite{Blagojevic:2000pi,Nester:2017wau}. However, for our practical purposes here this is unnecessary. As discussed
in the introduction \ref{introduction}, we could also obtain the equivalent dynamics by using the inertial variation that avoids
resorting to Lagrange multipliers \cite{Golovnev:2017dox}. This possibility was exploited in the (perturbative) calculation of  \cite{Conroy:2017yln}.},
\be \label{cccEoM}
\nabla_\mu\lambda_\alpha{}^{\mu\nu}= \nabla_\mu \lp \sqrt{-g}{P}^{\mu\nu}{}_\alpha\rp + \nabla_\mu \mathfrak{H}_\alpha{}^{\mu\nu}\,,
\ee
We have now noted that since $2\nabla_\mu \nabla_\rho\lambda_\alpha{}^{\nu\mu\rho} = [\nabla_\mu,\nabla_\rho]\lambda_\alpha{}^{\nu\mu\rho}=0$ where the first equality
is due to the antisymmetry of the Lagrange multiplier field and the second to the trivial curvature and torsion of the geometry. A similar argument will eliminate
$\nabla_\nu \nabla_\mu\lambda_\alpha{}^{\mu\nu} = 0$, and we can further assume that\footnote{This is 0) trivially true if there is no hypermomentum. Secondly, we
may consider that the 1) hypermomentum should be antisymmetric $\mathfrak{H}_\alpha{}^{(\mu\nu)}=0$ in which case our assumption is again identically true. Finally, was 
$\mathfrak{H}_\alpha{}^{(\mu\nu)} \neq 0$ we regard the statement as 2) the conservation law for the hypermomentum.} 
 $\nabla_\mu\nabla_\nu \mathfrak{H}_\alpha{}^{\mu\nu}=0$. The connection equation then boils down to
\be \label{cEoM}
\nabla_\mu\nabla_\nu \lp \sqrt{-g}{P}^{\mu\nu}{}_\alpha\rp =0\,.
\ee
On the other hand, the metric field equations can be written as
\be \label{qefe}
2\nabla_\alpha \lp\sqrt{-g}P^\alpha{}_{\mu\nu}\rp - q_{\mu\nu} -  \mL_Q g_{\mu\nu}=\sqrt{-g}\mathfrak{T}_{\mu\nu}\,,
\ee
where we have again used the torsion-free condition and defined
\be\label{def_quv}
q_{\mu\nu}\equiv 2\frac{\partial\mL_Q}{\partial g^{\mu\nu}}-\mL_Q g_{\mu\nu}\,,
\ee
for which, in the case (\ref{abcd2}), we obtain
\ba \frac{1}{\sqrt{-g}} q_{\mu\nu}
&=&c_1\Big(2Q_{\alpha\beta\mu}Q^{\alpha\beta}{}_\nu-Q_{\mu\alpha\beta}Q_\nu{}^{\alpha\beta}\Big)\nonumber\\
&+&c_2Q_{\alpha\beta\mu}Q^{\beta\alpha}{}_\nu+c_3\Big(2Q_\alpha Q^\alpha{}_{\mu\nu}-Q_\mu Q_\nu\Big)\nonumber\\
&+&c_4\tilde{Q}_\mu \tilde{Q}_\nu+c_5\tilde{Q}_\alpha Q^\alpha{}_{\mu\nu}\,.
\ea
Now using the general relation
\be \label{mder}
\nabla_\alpha\log\sqrt{-g}=-\frac12 g_{\mu\nu}\nabla_\alpha g^{\mu\nu}=\frac12 Q_\alpha\,,
\ee
we can obtain an explicit form of the field equations as
\ba
& c_1\Big[&(2\nabla_\alpha+Q_\alpha)Q^\alpha_{\phantom{\alpha}\mu\nu}  +   Q_{\mu\alpha\beta} Q_\nu^{\phantom{\nu}\alpha\beta}- 2Q_{\alpha\beta\mu}Q^{\alpha\beta}_{\phantom{\alpha\beta}\nu} \nn \\
& - &  \frac{1}{2}g_{\mu\nu}Q_{\alpha\beta\gamma}Q^{\alpha\beta\gamma}\Big]  + 
c_2 \Big[ (2\nabla_\alpha+Q_\alpha)Q^{\phantom{(\mu}\alpha}_{(\mu\phantom{\alpha}\nu)} 
\nn \\
 & - & Q_{\alpha\beta\mu}Q^{\beta\alpha}_{\phantom{\beta\alpha}\nu}  
  -   \frac{1}{2}g_{\mu\nu}Q_{\alpha\beta\gamma}Q^{\beta\alpha\gamma}\Big] +  
 c_3 \Big[  2\nabla_\alpha (g_{\mu\nu}Q^\alpha) \nn \\
 & + & Q_\mu Q_\nu - 2Q_{\alpha\mu\nu}Q^\alpha
  +  \frac{1}{2}g_{\mu\nu}Q_\alpha Q^\alpha\Big] + 
 c_ 4 \Big[ 2\nabla_{(\mu}\tilde{Q}_{\nu)} 
 \nn \\
 & + & Q_{(\mu}\tilde{Q}_{\nu)} - \tilde{Q}_\mu\tilde{Q}_\nu 
 -   \frac{1}{2}\tilde{Q}_\alpha\tilde{Q}^\alpha g_{\mu\nu} \Big] + 
 c_ 5 \Big[ \nabla_{(\mu}{Q}_{\nu)} \nn \\
 & + & \nabla_\alpha(g_{\mu\nu}\tilde{Q}^\alpha) 
 +  \frac{1}{2}Q_{\mu}{Q}_{\nu} - \tilde{Q}_\alpha{Q}^\alpha_{\phantom{\alpha}\mu\nu} \Big]  = \mathfrak{T}_{\mu\nu}\,.
\ea
Written in terms of the metric compatible covariant derivative in the torsionless spacetime, this becomes
\ba \label{qfe}
&c_1\Big[&   2\mathcal{D}_\alpha Q^\alpha_{\phantom{\alpha}\mu\nu} - 2Q_{\alpha\beta\mu}Q^{\beta\alpha}_{\phantom{\alpha\beta}\nu} + Q_{\mu\alpha\beta}Q_\nu^{\phantom{\nu}\alpha\beta}
\nn \\
& + &  2Q_{\alpha\beta(\mu}Q^{\phantom{\nu}\beta\alpha}_{\nu)} - \frac{1}{2}g_{\mu\nu}Q_{\alpha\beta\gamma}Q^{\alpha\beta\gamma}\Big]
\nn \\
& c_2  \Big[& 2\mathcal{D}_\alpha Q_{(\mu\nu)}^{\phantom{\mu\nu}\alpha} - Q_{\alpha\beta\mu}Q^{\alpha\beta}_{\phantom{\alpha\beta}\nu} 
+ Q_{\alpha\beta(\mu}Q_{\nu)}^{\phantom{\nu}\alpha\beta}
 \nn \\
 & + & Q_{\mu}^{\phantom{\mu}\alpha\beta}Q_{\nu\alpha\beta} 
  -  \frac{1}{2}g_{\mu\nu}Q_{\alpha\beta\gamma}Q^{\beta\alpha\gamma}\Big] + 
   \nn \\
  & c_3 \Big[& Q_\mu Q_\nu + g_{\mu\nu}\Big( 2\mathcal{D}_\alpha Q^\alpha 
 -  \frac{1}{2}Q_\alpha Q^\alpha\Big)\Big] + 
 \nn \\
 &c_4\Big[& 2\mathcal{D}_{(\mu} Q_{\nu)} - Q_\mu Q_\nu + Q_{(\mu} \tilde{Q}_{\nu)}
 \nn \\
 & - & Q_\alpha L^\alpha_{\phantom{\alpha}\mu\nu} -  \frac{1}{2}g_{\mu\nu}\tilde{Q}_\alpha\tilde{Q}^\alpha\Big] +
 \nn \\
 & c_5\Big[& \mathcal{D}_{(\mu} Q_{\nu)} - Q_\alpha L^\alpha_{\phantom{\alpha}\mu\nu} 
  +   \frac{1}{2}Q_\mu Q_\nu 
  \nn \\
  & + &g_{\mu\nu}\lp \mathcal{D}_\alpha\tilde{Q}^\alpha
    -  \frac{1}{2}{Q}_\alpha \tilde{Q}^\alpha\rp \Big] = \mathfrak{T}_{\mu\nu}\,.
\ea
As a cross-check, we insert the values (\ref{qgr}) and obtain 
\ba
\mathcal{D}_\alpha P^\alpha_{\phantom{\alpha}\mu\nu}  & + &   \frac{1}{2}\mathcal{D}_\mu Q_\nu + Q_{[\alpha\beta](\mu}Q^{\alpha\beta}_{\phantom{\alpha\beta}\nu)} -  \frac{1}{2}Q_\alpha P^\alpha_{\phantom{\alpha}\mu\nu} \nn \\
& - & \frac{1}{2}g_{\mu\nu}\lb \mathcal{Q} + \mathcal{D}_\alpha \lp Q^\alpha - \tilde{Q}^\alpha\rp \rb
\, \mathring{=}\,  -\mathfrak{T}_{\mu\nu}\,,\,\,\,\,\,
\ea
which, by comparison with (\ref{riccitensorq},\ref{ricciscalarq}), is equivalent, in the teleparallel spacetime, to the statement $\mathcal{R}_{\mu\nu} - \frac{1}{2}g_{\mu\nu}\mathcal{R} = \mathfrak{T}_{\mu\nu}$. 

Looking at the field equation (\ref{qefe}) we recognise the covariantly conserved term  
\be
2\nabla_\mu\lp \sqrt{-g}P^{\mu\nu}{}_\alpha\rp = \sqrt{-g}\lp 2\nabla_\mu + Q_\mu\rp P^{\mu\nu}{}_\alpha\,.
\ee
In the second equality we used again (\ref{mder}). Since the covariant divergence of this tensor density indeed vanishes according to (\ref{cEoM}),
and it is associated with the affine connection, we identify this tensor density with the inertial energy-momentum. 

\subsection{Symmetries} 
\label{symmetries}

In this section we will explore the symmetries of the theory from three perspectives: 
\begin{enumerate}
\item The possible existence of the so called $\lambda$-symmetries, which exist in the constraints sector of the action.
\item The gauge symmetries of the linearised theory, which depends on the form of the (quadratic) action containing the model-dependent kinetic terms.
\item The conservation of the energy-momentum,
which we would like to take for granted in any consistent theory as a consequence of the general coordinate invariance. 
\end{enumerate}

\subsubsection{Symmetries of the Lagrange multipliers}

The gauge symmetries of the Lagrange multipliers in the frame formulation of STG were checked by Adak {\it et al}, see e.g. \cite{Adak:2006rx}. As next, we shall have a closer look at the constraint sector of the Lagrangian density (\ref{abcd2}) that is given by
\be \label{lcons}
\mathcal{L}_\lambda =  \lambda_\alpha^{\phantom{\alpha}\beta\mu\nu} R^\alpha_{\phantom{\alpha}\beta\mu\nu} + \lambda_\alpha^{\phantom{\alpha}\mu\nu}T^\alpha_{\phantom{\alpha}\mu\nu}\,.
\ee
 Now, it is possible to introduce the transformations
\ba
\lambda_\alpha{}^{\beta\mu\nu} & \rightarrow & \lambda_\alpha{}^{\beta\mu\nu} + \lp \nabla_\rho + T_\rho\rp \kappa_\alpha{}^{\beta\mu\nu\rho} \nn \\
& + &  T^\mu{}_{\rho\lambda}\kappa_\alpha{}^{\beta\lambda\nu\rho} + \kappa_\alpha{}^{\beta\mu\nu}\,, \label{lambda3}\\
\lambda_\alpha{}^{\mu\nu} & \rightarrow & \lambda_\alpha{}^{\mu\nu}  + \lp \nabla_\beta + T_\beta\rp\kappa_\alpha{}^{\mu\nu\beta}\nn \\ & + & T^\mu{}_{\lambda\beta}\kappa_\alpha{}^{\lambda\nu\beta}\,,
\label{lambda4}
\ea
where both the $\kappa_\alpha{}^{\beta\mu\nu\rho}$ and the $\kappa_\alpha{}^{\beta\mu\nu}$ are tensor densities that are totally antisymmetric in their three last indices. They have thus 64 and 16 independent components,
respectively. By using the Bianchi identities (\ref{Bianchi1}-\ref{Bianchi3}) one can see that these redefinitions leave (\ref{lcons}) invariant, up to a total derivative. This means that the field equations
cannot completely determine the fields $\lambda_\alpha^{\phantom{\alpha}\beta\mu\nu}$ and $\lambda_\alpha^{\phantom{\alpha}\mu\nu}$, which have 96 and  24 independent components,
respectively. Let us notice however that, in the symmetric teleparallel theories considered in this section, the resolution of the Lagrange multipliers is not at all necessary and, therefore, the existence of undetermined components of the Lagrange multipliers is completely irrelevant.

\subsubsection{Linearised theory}

In order to gain some insight into the general quadratic theory, let us look at the perturbative degrees of freedom around a Minkowski metric background. We will fix the gauge so that the connection vanishes at all orders. It is easy to see that this is always possible by considering the general form of the connection imposed by the Lagrange multipliers. The vanishing of the curvature allows the connection to be only a pure $GL(4,\mathbb{R})$ transformation parametrised by $\Lambda^\alpha_{\phantom{\alpha}\mu}$ as in the metric teleparallel case discussed in the precedent section. The torsion-free condition further constrains the transformation matrix to satisfy $(\Lambda^{-1})^\alpha_{\phantom{\alpha}\nu}\partial_{[\mu} \Lambda^\nu_{\phantom{\nu}\beta]}=0$ so that the transformation can be parameterised as $\Lambda^\alpha_{\phantom{\alpha}\mu}=\partial_\mu\xi^{\alpha}$ with $\xi^\alpha$ some vector field. This is precisely how the trivial connection transforms under a general change of coordinates and, thus, we obtain the fundamental result of symmetric teleparallel theories that the connection can be exactly cancelled by a diffeomorphism\footnote{Which we will refer to as ''Diff'' in the following. Note that the active infinitesimal version of this transformation coincides with a translation in the tangent space, which would not affect the tangent space connection but transforms the spacetime affine connection (which is given
by the projected covariant derivative of the frame field).}. The gauge in which the connection is trivialised is called the {\it coincident gauge}.  In this gauge the connection can be parameterised by  $\xi^\alpha =M^\alpha{}_\beta x^\beta+\xi_0^\alpha$ with $M^\alpha{}_\beta$ an arbitrary constant (non-degenerate) matrix and $\xi_0$ a constant vector. Without loss of generality, we can transform the coordinates to have simply $\xi^\alpha=x^\alpha$, so we will take this as the vector parameterising the connection in the coincident gauge. If we now perform a linear Diff transformation in the coincident gauge we have $\xi^\alpha = x^\alpha + \delta \xi^\alpha$  while the metric transforms as
 $g_{\mu\nu} \rightarrow g_{\mu\nu} + 2\partial_{(\mu}\delta\xi_{\nu)}$. 
 
 Let the metric be perturbed around the flat background as $g_{\mu\nu}=\eta_{\mu\nu}+h_{\mu\nu}$, and let us fix the coincident gauge, $\Gamma^\alpha_{\phantom{\alpha}\mu\nu}=0$. Expanding the Lagrangian to the quadratic order in the perturbations gives then
\ba \label{h_action}
\mathcal{L}&=&c_1\partial_\alpha h_{\mu\nu} \partial^\alpha h^{\mu\nu}+(c_2+c_4)\partial_\alpha h_{\mu\nu}\partial^\mu h^{\alpha\nu}
\nonumber\\
&&+c_3\partial_\alpha h \partial^\alpha h+c_5\partial_\mu h^\mu{}_\nu\partial^\nu h\,.
\ea
As it is well-known, only a specific tuning of the coefficients in this Lagrangian describes a pure massless spin-2 field. One way to see this is by imposing the existence of some Bianchi identities that would be associated with additional gauge symmetries. Two possibilities arise, the first one being the usual linearised Diff invariance (that would be a symmetry additional to the original Diff symmetry that we used to remove the connection) corresponding to $c_2+c_4=-2c_1$, $c_3=-c_1$, $c_5=2c_1$, in accordance with the equivalent of GR given in (\ref{qgr}) up to the freedom in the choice of $c_4$ and the value of $c_1$ that can be fixed by the normalisation of $h_{\mu\nu}$. The other possibility is to impose the WTDiff symmetry, consisting of transverse diffeomorphisms plus a Weyl re-scaling, that corresponds to $c_2+c_4=-2c_1$, $c_3=-3c_1/8$, $c_5=c_1$. In both cases, the condition $c_2+c_4=-2c_1$ ensures the transverse gauge symmetry $\delta_{\zeta^\perp} h_{\mu\nu}=\partial_{(\mu}\zeta^\perp_{\nu)}$ with $\partial^\mu\zeta^\perp_\mu=0$. This symmetry is then completed either by unleashing the gauge parameter $\zeta_\mu$ so that it is no longer transverse or by adding the Weyl scaling $\delta_\phi h_{\mu\nu}=\phi\eta_{\mu\nu}$. Any other choice of the parameters will describe additional propagating degrees of freedom, which can then be associated to having dynamical degrees of freedom in $\xi^\alpha$. Notice that including $\xi^\alpha$ by working in an arbitrary gauge is equivalent to restoring the gauge symmetry of the quadratic action with $\xi^\alpha$ playing the role of the Stueckelberg fields and the parameters choices given above are those for which the quadratic action is oblivious to the Stueckelbergs.

\subsubsection{Conservation of the energy-momentum}

After clarifying the linearised theory, we shall then verify the consistency of the covariant conservation of energy-momentum non-perturbatively and for a generic theory. 
For this purpose, let us consider a generic action which may depend upon the $m$'th derivatives of $g_{\mu\nu}$, and define the variation wrt the metric 
\ba
2\sqrt{-g}\mathcal{G}_{\mu\nu} & \equiv & \frac{\delta \mL}{\delta g^{\mu\nu}} 
 =  \frac{\partial \mL}{\partial g^{\mu\nu}} - \partial_\alpha \frac{\partial \mL}{\partial g^{\mu\nu}{}_{,\alpha}} \nn \\
 & + & \dots + (-1)^m\partial_{\alpha_1\dots\alpha_m}\frac{\partial \mL}{\partial g^{\mu\nu}{}_{,\alpha_1\dots\alpha_m}} \,.
\ea
Thus (in contrast to the previously used $q_{\mu\nu}$ in (\ref{def_quv})), this variation $\mathcal{G}_{\mu\nu}$ is understood as the generalised Einstein tensor and it satisfies
\be
\mathcal{G}_{\mu\nu} = \mathfrak{T}_{\mu\nu}\,. 
\ee
What we would like to show is that $\nabla_{\mu}\mathcal{G}^{\mu\nu}=0$, which especially should hold also when choosing
the coincident gauge for the connection. We also generalise the definition of non-metricity conjugate as
\be \label{nmsuper2}
\mathcal{P}^\alpha{}_{\mu\nu}  \equiv \frac{\delta\mL}{\delta Q_\alpha{}^{\mu\nu}}\,.
\ee
We have simply denoted a generic conjugate density as $\mathcal{P}^\alpha{}_{\mu\nu}$ to emphasise that is now a density (just for convenience of the
present derivation) and that it needs not to be of the form (\ref{nmsuperquad}) (because we want our conclusion to depend only upon the symmetries
of the action and the chosen geometry). As previously, we can deduce from the equation of motion for the connection that
\be \label{eomc}
\nabla_\mu\nabla_\nu\mathcal{P}^{\mu\nu}{}_{\alpha}=0\,.  
\ee 
Considering the coordinate Diff $x^\alpha \rightarrow x^\alpha + \zeta^\alpha$, where $\zeta^\alpha$ is a vector vanishing on the boundary $\partial V$ of an $n$-dimensional region $V$, we require the invariance of the action over the volume of $V$,
\ba \label{diffa}
0 &=& \delta_{\zeta} S = \int_V \diff^n x \Big(\sqrt{-g} \mathcal{G}_{\mu\nu}\delta_{\zeta} g^{\mu\nu} + \frac{\delta \mL}{\delta \Gamma^\alpha{}_{\mu\nu}}\delta_{\zeta} \Gamma^\alpha{}_{\mu\nu} \nn \\
&+& R^\alpha{}_{\beta\mu\nu}\delta_{\zeta} \lambda_\alpha{}^{\beta\mu\nu} + T^\alpha{}_{\mu\nu}\delta_{\zeta} \lambda_\alpha{}^{\mu\nu}\Big)\,.
\ea
We have varied all the gravitational fields under the Diff: the metric, the connection, and the two Lagrange multipliers. The variations
of the latter vanish of course on shell, because the equations of motion for the multipliers just set the curvature and torsion to vanish. Further,
the equation of motion for the connection sets the second term to vanish identically.
It is however somewhat non-trivial that merely the covariant conservation (\ref{eomc}) of the inertial energy-momentum wrt the pure-gauge gravitational connection $\nabla$ now ensures the covariant conservation of the matter stress energy-momentum wrt the metric-compatible matter connection $\mathcal{D}$. This, in particular, means that we need to consider only the joint variation of the action wrt the metric and the non-metricity tensor, neither of which involves the Lagrange multipliers. 

To show this, we need the Diff transformation of each field, which is given by the Lie derivative of the respective field along the
vector $\zeta^\alpha$. For the metric it is well-known to be
\be \label{lieg}
\delta_{\zeta} g_{\mu\nu} = -2\mathcal{D}_{(\mu}\zeta_{\nu)}\,.
\ee
Starting from the transformation law of the affine connection, one can deduce that its Lie derivative is given as
\be \label{liec}
\delta_{\zeta} \Gamma^\alpha{}_{\mu\beta} = -\nabla_\mu\nabla_\beta \zeta^\alpha - \zeta^\lambda R^\alpha{}_{\beta\lambda\mu} -\nabla_\mu\lp \zeta^\lambda T^\alpha{}_{\lambda\mu}\rp\,.
\ee
We can then perform some partial integrations to obtain variation that is proportional to $\zeta^\alpha$ (and not its derivatives). For example,
for the first piece in (\ref{diffa}) we obtain, using (\ref{lieg}),
\ba
\int_V \diff^n x \sqrt{-g} \mathcal{G}_{\mu\nu}\mathcal{D}^{(\mu}\zeta^{\nu)} & = & \int_{\partial V} \sqrt{-g} \mathcal{G}_{\mu\nu}\zeta^\nu \diff (\partial V)^\mu \nn \\  -  \int_V \diff^n x\sqrt{-g}\mathcal{D}^\mu \mathcal{G}_{\mu\nu}\zeta^\nu&\,.&
\ea
The boundary term vanishes because of the assumption that $\zeta^\alpha=0$ at $\partial V$. Performing a couple of partial integrations for the
second term in (\ref{diffa}) as well, and substituting our variation of the connection which can now be written as
\be
\frac{\delta \mL}{\delta \Gamma^\alpha{}_{\mu\nu}} = \nabla_\rho\lambda_\alpha{}^{\nu\mu\rho}+\lambda_\alpha{}^{\mu\nu} - \mathcal{P}^{\mu\nu}{}_\alpha\,,
\ee
we find that variation of the action becomes (when dropping the multiplier terms from the above expression as it is easy to see they vanish identically),
\ba \label{diffb}
\delta_{\zeta} S & = & \int_V \diff^n x \Big( \sqrt{-g}\mathcal{D}^\mu \mathcal{G}_{\mu\lambda} + \nabla_\mu\nabla_\nu\mathcal{P}^{\mu\nu}{}_{\lambda}  \nn \\
& + & R^\alpha{}_{\nu\lambda\mu}\mathcal{P}^{\mu\nu}{}_\alpha - T^\alpha{}_{\lambda\nu}\nabla_\mu \mathcal{P}^{\mu\nu}{}_{\alpha}  \Big)\zeta^\lambda\,.
\ea
Since this $\delta_{\zeta} S =0$ for arbitary $\zeta^\lambda$, the expression in the parenthesis must disappear at each $x^\alpha \in V$. In fact, each of the four terms is independently zero.
One can note some conspiracies this required in STG. The second line in (\ref{diffb}) vanishes only because we have excluded torsion and curvature
from the geometry, and the conservation equation (\ref{eomc}) forces the second term in (\ref{diffb}) to vanish only because we have    
restricted the connection to be pure gauge. It is only then that the Diff of the affine connection (\ref{liec}) has the form mimicking the transformation leading to (\ref{eomc}).

To summarise, the theory will satisfy the generalised Bianchi identity, which amounts now to the vanishing of the metric-divergence of the right hand side of the equation (\ref{qefe}). That is crucial for the consistency with minimally coupled matter on the left hand side that then obeys $\mathcal{D}_\mu \mathfrak{T}^\mu_{\phantom{\mu}\nu}=0$.  The identity is a consequence of the construction of the action (\ref{abcd}) to be a scalar under the Diff, and therefore the same conclusion holds for any invariant action. Note that we can always transform to the coincident gauge, where the connection field still presents an equation of motion or, in other words, the connection field equations do not trivialise in the coincident gauge. The situation is analogous with the usual teleparallel torsion theories, where though the inertial connection could be set to zero by a suitable Lorentz transformation, the additional degrees of freedom decouple from the theory only when the action supports an additional Lorentz symmetry so that the spin connection identically disappears from the equations of motion \cite{Tamanini:2012hg,Krssak:2015rqa,Golovnev:2017dox,Krssak:2017nlv}. Otherwise, the tetrad
field equations assume antisymmetric components. These are redundant with the spin connection equations, so the unitary gauge is restrictive but legitimate. In the case of STG here, we find that the connection equation of motion is redundant with the energy-momentum conservation of matter, so the consistency of the solutions in the unitary gauge is guaranteed by imposing that $\mathcal{D}_\mu \mathfrak{T}^\mu_{\phantom{\mu}\nu}=0$. The unitary gauge of a pure-gauge theory is briefly discussed in the more trivial case of electromagnetism in the Appendix \ref{sec:em}.  

It is interesting to note in passing that the vector $\xi^\alpha$ is related to the radius field \cite{Hehl:1994ue}.
We recall that a tangent space coordinate vector enters into the theory in its gauge formulation via the relation between the translation gauge potential and the tetrad to reconcile their
property of transforming as a tensor and as a connection, respectively. It is a manifestation of the generator of the inhomogeneity in Cartan's geometry, a vector whose joint dynamics with the gauge potentials of the relevant (pseudo-)orthogonal symmetries have been recently illustrated with the idealised waywiser \cite{Westman:2012xk,Westman:2014yca}. We have found that in STG, essentially the same vector field reappears as the St\"uckelberg of the translation symmetry that singles out GR from the generic quadratic theory. One can interpret that when $\xi^\alpha=x^\alpha$, the origins of the tangent space and the spacetime coincide. As the more usual spin connection describes the
local rotation wrt the tangent space its local displacement is described by the translation connection. 

\section{Applications}
\label{sec:applications}

In this section we illustrate some aspects of the theories in STG with examples. 
The cosmological equations are considered in \ref{sec:cosmology}, to see some basic properties of the system. 
Their Newtonian limit is briefly checked in \ref{sec:flat}. The self-coupling of the linear theory, leading to its non-linear completion now in a perhaps more unique and
consistent fashion, is commented in \ref{sec:strap}, and finally the black hole entropy is deduced from a regular Euclidean action in \ref{sec:euclidean}.

\subsection{Cosmology}
\label{sec:cosmology}

As an example, we consider the homogeneous and rotation-invariant, but time-dependent solutions
that describe the cosmological expansion by the line element
\be \label{frw}
\diff s^2 = -N^2(t)\diff t^2 + a^2(t)\delta_{ij}\diff x^i \diff x^j\,,
\ee
where $a(t)$ is the scale factor, the expansion rate is defined as $H=\dot{a}/a$. However, now it is crucial to include the lapse function $N(t)$. As we saw above in Section \ref{symmetries}, in the generic theory the diffeomorphism gauge
symmetry is lost if we pick the coincident gauge $\Gamma^\alpha_{\phantom{\alpha}\mu\nu}=0$, and therefore we cannot always choose the time parameterisation arbitrarily
as we are used to in GR cosmology\footnote{On the other hand, we could keep this freedom to reparameterise the metric if we allowed for an non-vanishing inertial connection.
The most general vector compatible with homogeneity and isotropy includes two time-dependent functions $\xi^0(t)$, $\xi(t)$, as $\xi^\alpha=(\xi^0(t),\xi(t),\xi(t),\xi(t))$.}. 

Let us know compute the relevant objects for the ansatz (\ref{frw}) and in the coincident gauge so $\Gamma^\alpha_{\mu\nu}=0$ and, thus, covariant derivatives are simply partial derivatives. For the non-metricity we have $Q_{\alpha\mu\nu}=\delta^0_\alpha \dot{g}_{\mu\nu}$ so the non-vanishing components are
\be \label{Qfrw}
Q_{0 ij} = 2H g_{ij}\,, \quad Q_{0 00} = -2\dot{N}N
\ee
and the rest of the components vanish identically,
\be
Q_{i\mu\nu} = Q_{\mu 0 i} = Q_{\mu i 0} = 0\,.
\ee
The nonzero connection coefficients in (\ref{disformation2a}) then are 
\be \label{lfrw}
L^0{}_{00}=-\frac{\dot{N}}{N}\, ,\quad L^0_{\phantom{0}ij}   =  -\frac{H}{N^2} g_{ij}\,, \quad L^i_{\phantom{0}0j}  = L^i_{\phantom{0}j0} = -H\delta^i_j\,, 
\ee
and in (\ref{disformation2b}),
\ba \label{hatlfrw}
\hat{L}^0_{\phantom{0}00} &=&-3H-\frac{\dot{N}}{N}\,, \quad
 \hat{L}^0_{\phantom{0}ij}   =-\frac{1}{N^2}  \left(3H+\frac{\dot{N}}{N}\right)g_{ij}\,, \nonumber \\
 \hat{L}^i_{\phantom{0}0j}  & = &  \hat{L}^i_{\phantom{0}j0}  = -\left(3H+\frac{\dot{N}}{N}\right)\delta^i_j\,.
\ea
As it should in the gauge we are in, $L^\alpha_{\phantom{\alpha}\mu\nu} = -\left\{^{\phantom{i} \alpha}_{\mu\nu}\right\}$. Since from (\ref{lfrw},\ref{hatlfrw}) we have in (\ref{qdef}) that $L^0_{\phantom{0}ij}-\hat{L}^0_{\phantom{0}ij} =(2H+\dot{N}/N)/N^2g_{ij}$, we obtain 
by using this together with (\ref{Qfrw}) that $\mQ=6H^2/N^2$ as it corresponds for the Lagrangian the GR equivalent. In the general quadratic case, the mini-superspace action takes the form 
\begin{align}
\mS=\int\diff t\frac{a^3}{N}\left[6\big(c_1+3c_3\big)\frac{\dot{a}^2}{a^2}+2\tilde{c}\frac{\dot{N}^2}{N^2}+6\big(2c_3+c_5\big)\frac{\dot{a}\dot{N}}{aN}\right]
\end{align}
where $\tilde{c}=\sum_{i=1}^5 c_i$. This action clearly shows that the lapse is indeed a dynamical degree of freedom unless $\tilde{c}=0$ and $2c_3+c_5=0$, confirming the necessity to keep it for the consistency of the cosmological equations. The two independent field equations can be computed by taking variations in the above action or directly from the covariant field equations (\ref{qfe}), resulting in the generalised Friedmann equations given by
\begin{align}
&6\lp c_1 + 9c_3 + 3c_5\rp H^2 + 6\lp 2c_3 + c_5 \rp\dot{H}  \\
&+ \tilde{c} \left[4\frac{\diff}{\diff t}\left(\frac{\dot{N}}{N}\right)-2\left(\frac{\dot{N}}{N}\right)^2+12H \frac{\dot{N}}{N}\right]=N^2\rho\, ,\nn\\
& 2(c_1+3c_3)\left[3H^2+2\dot{H}-2H\frac{\dot{N}}{N}\right]
+ 2(2c_3+c_5)\frac{\diff}{\diff t}\left(\frac{\dot{N}}{N}\right) \nn \\  
&-  2(\tilde{c}+2c_3+c_5)\left(\frac{\dot{N}}{N}\right)^2=-N^2p\,,
\end{align}
where we have set $8\pi G=1$ and included a perfect fluid in the matter sector. These equations show again the dynamical nature of the lapse that cannot be removed by a simple time reparameterisation as in the usual GR case. The perfect fluid source, with the energy density $\rho$ and the pressure $p$, is subject to the continuity equation $\dot{\rho} + 3H(\rho + p) =0$, which leads to an equation that is independent of the matter source and, furthermore, is equivalent to the connection equation of motion in the coincident gauge. This equation reads 
\ba
0 & = & 3\lp 2c_3+c_5\rp\lp \ddot{H}+9H\dot{H}+9H^3\rp  \\
& + & 18\lp \tilde{c}-2c_3-c_5\rp H^2 \left(\frac{\dot{N}}{N}\right) \nn \\
& + & 3\lp 4\tilde{c}-2c_3-c_5\rp \left[H \frac{\diff}{\diff t}\left(\frac{\dot{N}}{N}\right) -H \left(\frac{\dot{N}}{N}\right)^2\right] \nn \\
& + & 2\tilde{c}\left[ \frac{\diff^2}{\diff t^2}\left(\frac{\dot{N}}{N}\right) 
-  3\left(\frac{\dot{N}}{N}\right)\frac{\diff}{\diff t}\left(\frac{\dot{N}}{N}\right) + \left(\frac{\dot{N}}{N}\right)^3\right]\,.\nn \label{extra}
\ea
Since this equation is redundant with the connection equation, we can easily check that it completely trivialises for the parameters corresponding to the GR equivalent. In fact, we can readily see that it trivialises for the more general theories with $\tilde{c}=0$ and $2c_3+c_5=0$, that correspond to those theories for which the dynamical part of the lapse in the mini-superspace action completely drops. We can thus  associate this identity to time-reparameterisations invariance.

For the sake of illustration, let us check the particular solutions with $N=1$, i.e., we seek for cosmological solutions where the time coordinate actually corresponds to the usual cosmic time. The Friedmann equations are then considerably simplified, and read\footnote{An equivalent (slightly more general) parameterisation of the Friedmann equations was studied inspired by entropic cosmology \cite{Koivisto:2010tb}.
In Ref. \cite{Adak:2008gd}, this pair of equations was derived for the symmetric teleparallel theory, and an expansion driven by a modified Chaplygin gas equation of state was shown to arise as a solution. However, in our interpretation this is not entirely correct because the system (\ref{f1},\ref{f2}) would be non-conservative. 
Constraints from particle experiments in a non-conservative theory are currently considered in Ref. \cite{Latorre:2017uve}, but there it should take into account that the non-minimal 
couplings are an artifact of a transformation to the Einstein frame. Of course, in principle a non-conservative version of the theory could also be viable, then within more stringent experimental limits \cite{ Koivisto:2010tb,Latorre:2017uve}.}
\ba 
6\lp c_1 + 9c_3 + 3c_5\rp H^2 + 6\lp 2c_3 + c_5 \rp\dot{H} & = & \rho\,,\,\,\, \label{f1} \\
-2(c_1+3c_3)(3H^2+2\dot{H}) & = & p\,.\,\,\,\,\,\,\,\,\,\, \label{f2}
\ea
Using the time derivative of the first equation, it is quick to verify that in this case the continuity $\dot{\rho} + 3H(\rho + p) =0$  requires that $\ddot{H}+9\dot{H}H+9H^3=0$, unless
$2c_3=-c_5$. This is indeed the  
first line in the equation (\ref{extra}), which is the only one surviving the limit $N=1$. The system is then, however, over-determined and can only be compatible with special kind of matter.
If $2c_3=-c_5$, the Friedmann equations reduce to the usual ones, up to a redefinition of the
Newton's constant. We see that the cosmology of the quadratic models, when restricting to solutions without evolving lapse function, allows only viable GR-like solutions. To find more general dynamics
in the quadratic models, one needed to take into account the dynamics of the lapse. It can be interesting to explore cosmologies with dynamical lapse in the quadratic and more general models, since 
such models introduce the time-dilation rate on a similar physical footing as the space-expansion rate, their effect to the redshift being degenerate. Perhaps in the presence of two expansion rates one could implement an anamorphic universe \cite{Ijjas:2015zma}.   

We will know skim the cosmology of some more general theories beyond the quadratic class considered so far. Among them, the $f(\mQ)$ models (which are the analogues of $f(\mathcal{R})$ and $f(\mathring{\mathbb{T}})$ theories) are special because the connection equation vanishes
identically in a FLRW background and, consequently, the field equations in the coincident gauge imply conservation in the matter sector identically. In other words, the metric field equations directly entails the continuity equation $\dot{\rho}+3H(\rho+p)=0$ and non-trivial dynamics are available requiring an evolving lapse function.  Note that for the coincident GR, which is the quadratic case with $f(\mQ) \sim \mQ$, the connection equation vanishes not only for cosmological solutions but for all backgrounds. This is of course a direct consequence of the full gauge symmetry present in coincident GR. Since the symmetry is realised only up to total derivative terms, it is expected that its functional extensions as e.g. $f(\mQ)$ will lose the symmetries. However, we have seen here that a cosmological background retains some symmetry for the $f(\mQ)$ extensions which in turn is responsible for the trivialisation of the connection equation. The presence of such a symmetry can be nicely identified in the corresponding action in the mini-superspace
\be
\mS=\int\diff^4x\sqrt{-g}f(\mQ)=\int\diff^3x\diff t Na^3f(6H^2/N^2).
\ee
It is then straightforward to see that this action features a time reparameterisation symmetry $t\to\zeta(t)$,  $N(t)\to N(t)/\dot{\zeta}(t)$. Several exact background cosmological solutions of the $f(\mQ)$ models  were already presented in \cite{BeltranJimenez:2017tkd}. 
Further, it turned out that the $f(\mQ)$ background cosmologies are
equivalent to those in the extensively studied $f(\mathring{\mathbb{T}})$ models \cite{Cai:2015emx,Bahamonde:2016grb,Hohmann:2017jao}, and amongst many
other classes of modified gravity cosmologies, they are captured by the ''vector distortion'' parameterisation \cite{Jimenez:2016opp,Jimenez:2015fva}. The fact that the cosmology in the $f(\mQ)$ class is equivalent to that found in the $f(\mathring{\mathbb{T}})$ theories is a non-trivial result since, as we discussed above, the corresponding equivalents to GR differ by a boundary term. It happens however that such a boundary term vanishes for a FLRW metric. It is however an interesting issue that merits further study, that the perturbation evolution could distinguish between these models.
Of course, more general symmetric teleparallel models, e.g. defined by a nonlinear function $f(Q_{\alpha\beta\gamma}P^{\alpha\beta\gamma})$ can be interesting to study, in analogy to the $f(\mathbb{T})$ models recently 
discussed in \cite{Bahamonde:2017wwk}.

Let us finish this brief discussion on cosmological applications by considering some perturbations. As usual, the simplest modes to analyse around a cosmological background are the tensor modes so let us start by studying those. We will stick to the coincident (connection-free) gauge so that $\Gamma^\alpha_{\mu\nu}=0$ also at the perturbative level. On the other hand, tensor modes are gauge-invariant and we can parameterise them in the usual way 
$\delta g_{ij}=-a^2 h_{ij}$, with $\delta^{ij}h_{ij}=\partial^i h_{ij}=0$. With these definitions, the equations for the tensor perturbations can be written as
\be \label{gw}
\ddot{h}_{ij}+\left(3H+\frac{\dot{N}}{N}\right)\dot{h}_{ij}-\frac{N^2}{a^2}\nabla^2h_{ij}
=-\frac{N^2}{2c_1}\Pi_{ij}
\ee
with $\Pi_{ij}$ the transverse and traceless part of the energy-momentum tensor. Let us notice that for $c_1=-1/4$ we obtain the GR equation for the propagation of gravitational waves.
In the case of the coincident GR, we can reparameterise the metric freely to get rid of the $N$-term. Recall that in more general theories however, $N$ is a dynamical degree of freedom and 
therefore the speed of gravitational waves is effectively modified according to this equation. Nevertheless, the gravitational wave travels with the speed of light. It is important to see that the propagation of photons will exhibit the same effective $N^2$-modulation. It can be interesting to study whether this conclusion holds in more general backgrounds.

\subsection{Flat background}
\label{sec:flat}

Now we can turn to the scalar modes. In order to simplify the analysis we will consider a Minkowski background, that should coincide with the result for sub-Hubble modes. We will parameterise the perturbations as
\be
\delta g_{00}=-2\phi,\quad\delta g_{0i}=\partial_i B,\quad \delta g_{ij}=-2\psi\delta_{ij}+\partial_i\partial_j E\,.
\ee
The metric field equations for the general quadratic theory are then
\ba
&&-6(2c_3+c_5)\psi_k''-2(6c_3+c_5)k^2\psi_k\\
&&+4(c_1+c_2+c_3+c_4+c_5)\phi_k''+4(c_1+c_3)k^2\phi_k\nonumber\\
&&-2(c_2+c_4+c_5)k^2B_k'-(2c_3+c_5)\big(E_k''+k^2E_k\big)=\delta T^{00}\,,\nonumber\\
\nonumber\\
&&2(c_2+c_4+3c_5)\psi_k'-2(c_2+c_4+c_5)\phi_k'\\
&&(2c_1+c_2+c_4)\big(B_k''+k^2B_k\big)+(c_2+c_4+c_5)k^2E_k'\nonumber\\
&&=-\frac{i}{k^2}k_i\delta T^{0i}\,,\\ \nonumber
\ea
\ba
&&12(c_1+3c_3)\psi_k''+4(3c_1+c_2+9c_3+c_4+3c_5)k^2\psi_k\\
&&-6(2c_3+c_5)\phi_k''-2(6c_3+c_5)k^2\phi_k\nonumber\\
&&+2(c_2+c_4+3c_5)k^2B_k'\nonumber\\
&&+2k^2\left[(c_1+3c_3)E_k''+(c_1+c_2+3c_3+c_4+2c_5)k^2 E_k\right]\nonumber\\
&&=\delta_{ij}\delta T^{ij}\,,\nonumber\\
\nonumber\\
&&2(2c_2+2c_4+3c_5)\psi_k-2c_5\phi_k+2(c_2+c_4)B_k' \\
&&2c_1 E_k''+(2c_1+2c_2+2c_4+c_5)k^2E_k=\sigma\,,\nonumber
\ea
For completeness we have included the perturbative matter sources in the right hand side, $\sigma$ being the scalar potential of the anisotropic stress
that appeared in (\ref{gw}).
It is instructive to check the equations for the GR values of the parameters:
\ba
2k^2\psi_k&=&-\delta T^{00}\,,\\
2\psi_k'&=&\frac{i}{k^2}k_i\delta T^{0i}\,,\quad\quad \\
6\psi_k''+k^2\left[E_k''-2(\phi_k-\psi_k+B_k')\right]&=&\delta_{ij}\delta T^{ij}\,,\quad\\
k^2\left[E_k''-2(\phi_k-\psi_k+B_k')\right]&=&-2\sigma\,.
\ea
From the general equations we can obtain the number of propagating scalar modes by computing the corresponding dispersion relation, which in this case is given by
\be
c_1\kappa_1\kappa_2\left(\omega^2-k^2\right)^4=0\,,
\ee
with
\ba
\kappa_1 &= &2c_1+c_2+c_4\,,\\
\kappa_2 &= & 4c_1^2+12c_3(c_2+c_4)-3c_5^2 \nonumber \\ &+ & 4c_1(c_2+4c_3+c_4+c_5)\,.
\ea
We thus see that either we have four scalar modes propagating at the speed of light (as expected since we are not breaking Lorentz invariance) or the dispersion relation trivialises. It is not difficult to see that this is the case for the symmetric teleparallel equivalent of GR values (\ref{qgr}), but there are other possibilities.

\subsection{Couplings of gravity in STG}

We would like to point out two theoretical advantages of STG, compared to GR and to TEGR, respectively. They are:
\begin{itemize}
\item The invariant $\mathcal{Q}$ can be bootstrapped.   
\item The minimal coupling of spinors is viable. 
\end{itemize}
In the rest of this subsection we explain these in more detail.

\subsubsection{The self-coupling}
\label{sec:strap}

If one wants to consider a quadratic form for the symmetric rank-2 tensor $h_{\mu\nu}$, a Lorentz metric $\eta^{\mu\nu}$ is implied. 
The quadratic action which is invariant with respect to the independent transformations of the tensor is
\ba \label{mass} 
L_0 = &   - &\frac{1}{2} \nabla_\alpha h_{\mu\nu}\Big[( \eta^{\alpha\beta}\eta^{\alpha\beta}\eta^{\mu(\nu}\eta^{\rho)\sigma} 
\nn \\ & + &  2\eta^{\alpha\sigma}\eta^{\nu(\rho}\eta^{\nu)\beta} )   \Big] \nabla_\beta h_{\rho\sigma} + \lambda h_{\mu\nu}\tau^{\mu\nu}\,.
\ea
The last term represents sources, with a coupling constant $\lambda$. Especially, the gravitational field is taken to couple to its 
own energy-momentum. This is now naturally given by the variation wrt the metric $\eta^{\mu\nu}$. In the symmetric teleparallel paradigm, the geometry is non-orthononormal and the gravitational interaction, like the interactions  in Yang-Mills theory, are described as internal geometry. The $\eta^{\mu\nu}$ is not a constant in STG. The prescription for any field $\phi$ representing the algebra of $\eta^{\mu\nu}$ with a lagrangian $L(\phi,\nabla\phi)$ is 
\be \label{emt}
\tau_{\mu\nu} =- \frac{2}{\sqrt{-\eta}} \frac{\partial (\sqrt{-\eta}L)}{\partial\eta^{\mu\nu}}\,.
\ee
In the present case of $L=L_0$, we obtain an expression of the form
\be
\tau^{\mu\nu} = K^{\mu\nu\kappa\lambda\alpha\beta\rho\sigma}(\nabla_\alpha h_{\kappa\lambda})(\nabla_\beta h_{\rho\sigma})\,, 
\ee
with an awkward tensor $K$, constructed with only the $\eta^{\mu\nu}$. The formula $\tau^{\kappa\lambda}$ should reproduce the Tolman's expression for the (pseudo-tensor of) gravitational energy-momentum at the first non-trivial order. 

One can deduce the full non-linear theory of any field $\phi$ coupled to gravitation using the prescription (\ref{emt}). This can be easily seen when taking advantage of the previous insights into the bootstrap \cite{Deser:1969wk,Deser:2017ihr,Tomboulis:2017fim} that were clarified in Ref. \cite{Padmanabhan:2004xk}. 
The idea is that the gauge symmetry dictates the non-linear self-couplings as well as it specifies the quadratic form.
Consider the second-rank symmetric tensor $h_{\mu\nu}$ and take as the starting point its unique quadratic form, in our case (\ref{mass}). The tensor is self-coupled with
the source $\tau^{\mu\nu}_0$ given by (\ref{emt}) where now $L=L_0$. We can then regard $L_1 = L_0 + \frac12\lambda h_{\mu\nu}\tau^{\mu\nu}_0$ as the second-order approximation to the theory. 
Again we have to take into account its self-coupling, and now the source term $\tau^{\mu\nu}_1$ is given by the variation (\ref{emt}) with $L=L_1$. We obtain the next approximation,
$L_2=L_1 + \frac{1}{3!}\lambda h_{\mu\nu}\tau^{\mu\nu}_1$,  then let it self-couple to get the $L_3$, and from this the $L_4$, and so on until $L_\infty$. The computation is formally the same as in Ref. \cite{Padmanabhan:2004xk}, and as was clarified there, the result is rather the Einstein than the Einstein-Hilbert action.
We now obtain
\be \label{L_infty}
L_\infty = \frac{1}{\lambda^2} \mathcal{Q} \,,
\ee
which is the covariantised version of the Einstein action\footnote{By this we mean the Lagrangian
\be \label{gg}
\mL_{\rm Einstein} = g^{\mu\nu}\Big(\left\{^{\phantom{i} \alpha}_{\beta\mu}\right\} \left\{^{\phantom{i} \beta}_{\nu\alpha}\right\} -\left\{^{\phantom{i} \alpha}_{\beta\alpha}\right\}\left\{^{\phantom{i} \beta}_{\mu\nu}\right\}\, \Big)\,,
\ee
which differs from the Einstein-Hilbert Lagrangian by a boundary term, $\mathcal{R}=\mL_{\rm Einstein} + \mL_{\rm B}$, where the boundary term (a total derivative) can be written as
\be \label{bb}
\mL_{\rm B} = g^{\alpha\mu}\mathcal{D}_\alpha \left\{^{\phantom{i} \nu}_{\mu\nu}\right\} - g^{\mu\nu}\mathcal{D}_\alpha \left\{^{\phantom{i} \alpha}_{\mu\nu}\right\}\,. 
\ee
By noticing that the non-metricity scalar can be expressed in terms of the disformation as
\be \label{q}
\mQ = g^{\mu\nu}\lp L^\alpha{}_{\beta\mu}L^\beta{}_{\nu\alpha} - L^\alpha{}_{\beta\alpha}L^\beta{}_{\mu\nu}\rp\,,
\ee
and using that the coincident gauge gives the relation $L^\alpha_{\mu\nu}=-\left\{^{\phantom{i} \alpha}_{\mu\nu}\right\}$, one can readily see that the coincident GR Lagrangian is by construction equivalent to the Einstein Lagrangian (\ref{gg}), i.e., when the covariant derivative reduces to the partial one in the coincident gauge we have
\be
\nabla_\alpha \overset{0}{=} \partial_\alpha\,, \quad  \mathcal{Q} \overset{0}{=} \mL_{\rm Einstein}\,.
\ee 
}.
Tomboulis has also recently presented a covariantised version of the Einstein action that was based on the introduction of an additional reference metric \cite{Tomboulis:2017fim}, and shown to clarify how the different definitions of energy-momentum pseudo-tensors in the literature are related by superpotentials. On the other hand, the possibililty of covariantly defined energy-momentum has been established in teleparallelism \cite{moller}, and found, in the symmetric teleparallel formulation, to reproduce Tolmans definition \cite{Nester:1998mp}. At least, the formulation in STG has the technical
benefit of retaining the simplicity of the Einstein action, which perhaps is less the case in the bi-metric construction with a reference metric. Conceptually, the result (\ref{L_infty}) emerges now more uniquely and without the assumptions of additional fields. Firstly, our general linear first principle has promoted the partial derivatives to the gauge-covariant derivatives, and each
of the infinite steps in the bootstrap manifestly respects the symmetry. Secondly, as explained above, the prescription (\ref{emt}) is the natural consequence of the symmetric teleparallel paradigm.
 
\subsubsection{The matter coupling}
\label{spinor}

In discussing hypermomentum in Section \ref{telesec}, we mentioned that the coupling of matter can be somewhat ambiguous in generalised geometries. In the torsion teleparallelism, the most straightforward
implementation of a minimal coupling principle to fermions fails \cite{Obukhov:2002tm}. Using the Weitzenb\"ock connection in the Dirac Lagrangian does not yield a consistent theory. The resulting canonical
energy momentum tensor for the Dirac field would be asymmetric even if the geometric part of the gravitational field equation is symmetric. 
Various viewpoints to this issue have been presented in the literature \cite{Mosna:2003rx,Maluf:2003fs,Mielke:2004gg,Obukhov:2004hv}. It has been pointed out that 
the coupling of spin can be made consistent either by the change of the coupling prescription or by the change of the dynamical scheme \cite{Obukhov:2004hv}. While the second
alternative would lead to the framework of Einstein-Cartan theory, the first option is the standard prescription used in TEGR \cite{Hayashi:1979qx,Aldrovandi:2013wha,Maluf:2013gaa}. 
There, one exploits a coupling which is formally equivalent to GR, i.e. instead of the Weitzenb\"ock connection one plugs the Riemannian Levi-Civita connection (\ref{christoffel}) into the Dirac Lagrangian.
Note, that this promotes the framework into the hybrid geometry reminiscent of the hybrid metric-Palatini mentioned in the footnote \ref{hybrid}. This obviously renders the matter sector of the theory viable, 
since it is by construction precisely the same as in GR. One may avoid referring to the Levi-Civita connection (\ref{christoffel}) explicitly, since in the Weitzenb\"ock geometry with a vanishing spin connection, 
one can re-express the Levi-Civita connection (\ref{christoffel}) as minus the contortion (\ref{contortion}), but this renaming of the variables is only a cosmetic improvement in the sense that one nevertheless
 resorts to a different connection, introduced arguably ad hoc in order to deal with spinning matter.
 
In STG it is not necessary to consider any adjustments to the minimal prescription to consistently incorporate Dirac matter. That is, the derivatives $\partial_\alpha$ of the usual flat-space Dirac Lagrangian 
can be understood as $\nabla_\alpha$ (in the coincident gauge), and this remains valid when gravitation is turned on. It turns out that though in the action we use the covariant derivative $\nabla_\alpha$, 
in the equations of motion for the Dirac field there instead appears the $\mathcal{D}_\alpha$. Spinors naturally couple only to the metric part of the connection. This is due to the Hermitean property of the 
Dirac action (which ensures that the equation of motion for the conjugate spinor is the conjugate of the equation of motion for the spinor). In the coincident gauge we have $\Gamma^\alpha{}_{\mu\nu} =
\left\{^{\phantom{i} \alpha}_{\mu\nu}\right\} + L^\alpha{}_{\mu\nu} = 0$, from which the Hermitean action neatly filters out the piece $L^\alpha{}_{\mu\nu}$, and thus predicts the standard equation of motion
for the spinors in curved spacetime, involving only the $\left\{^{\phantom{i} \alpha}_{\mu\nu}\right\}$ \cite{Koivisto:2018aip}. The Dirac equation in STG had been derived, in the language of exterior forms, also in Ref. \cite{Adak:2008gd}, where essentially the piece $\left\{^{\phantom{i} \alpha}_{\mu\nu}\right\} $ was re-interpreted as $-L^\alpha{}_{\mu\nu}$.  In addition, the derivation on a spinor bundle
can reveal also a new complex partner of $Q_\alpha$ that may have relevance in the unification of electromagnetism into geometry \cite{Koivisto:2018aip}. 
It is well-known that Dirac Lagrangian indeed filters out the scale connection piece $\hat{L}^\alpha{}_{\mu\nu}$ due
to a Weyl non-metricity $Q_\alpha$ \cite{Blagojevic:2002du}, but it has not been often clarified that in fact the Dirac fields are oblivious to any disformation of geometry, described by a general  $Q_{\alpha\mu\nu}$. However, the same conclusion was arrived at in an index-free formalism \cite{Formiga:2012ns}. 

The analogous procedure of using the $\nabla_\alpha$ in the Dirac Lagrangian does not work in the TEGR context, since in addition to the Levi-Civita connection, the (totally antisymmetric component of the) torsion couples with the Dirac fields, and for the Weitzenb\"ock connection they cancel each other, leaving no geometrical effects for spinors to account for gravity.
On the other hand, one could also use the standard TEGR way of dealing with spinors in the STG, that is, replace the Levi-Civita connection into the equation of motion for the spinors, rewritten in terms of non-metricity, i.e. employ a connection $\Gamma^\alpha{}_{\mu\nu} = -L^\alpha{}_{\mu\nu}$ (such a prescription was deduced for the parallel transport of vectors in STG \cite{Adak:2011rr}), then arriving at the
Dirac equation in the from presented in Ref. \cite{Adak:2008gd}. However, the desired result follows simply by $\partial_\alpha \rightarrow \nabla_\alpha$ in the STG Dirac action. It remains to investigate non-minimal couplings of spinors in the STG.

\subsection{The double copy}

In this subsection we discuss how some interesting developments in gravity theory might be understood in the framework of the coincident GR. In particular, we comment on its relationship to some previous studies of alternative variational principles, and on the possible relevance to the double-copy structure of the gravity scattering amplitudes.

\subsubsection{Two metrics and variational principles}

In the linear analysis, we noticed that the action possesses a double symmetry. The combination $h_{\mu\nu}$ is invariant, but furthermore the action of coincident
GR happens to be symmetric also under the diffeomorphism-like transformation $h_{\mu\nu} \rightarrow h_{\mu\nu} + 2\zeta_{(\mu,\nu)}$. This double symmetry is 
retained in the non-perturbative action, since firstly, the $\mathcal{Q}$ is covariant, and secondly, restricting to the coincident gauge, it is invariant up to a boundary term.
This boils down to that the theory is symmetric under both the independent translations of the metric and the indepedent translations of the connection. 

Being the theory dynamically equivalent to GR, one would expect the doubled symmetry to be somehow present there as well. Indeed, the application of the bimetric
variational principle \cite{Koivisto:2011vq} reveals that there there is a pair of (massless) Fierz-Pauli terms hiding in the Einstein-Hilbert 
action \cite{BeltranJimenez:2012sz}. 

To explain this, we recall the C-theory \cite{Amendola:2010bk}, where the connection is generated by a metric
$\hat{g}_{\mu\nu}$ that may have a non-trivial, possibly even a curvature-dependent, conformal relation to the metric $g_{\mu\nu}$, such as 
$\hat{g}_{\mu\nu} = C(g^{\mu\nu}\hat{R}_{\mu\nu})g_{\mu\nu}$. In the prototype $f(g^{\mu\nu}\hat{R}_{\mu\nu})$ models \cite{Capozziello:2015lza}, we then reproduce
the Palatini version in the limit $C=f'$ and the metric version in the limit $C=1$. We notice that these coincide in the case of the Einstein-Hilbert action. In passing we note that is however yet a simpler $C$-theory, the $C=0$ theory, which could provide an alternative means to trivialise the affine geometry, and to covariantise the Einstein action (for which reference metrics have been considered previously \cite{Tomboulis:2017fim}, and we also recall, from footnote \ref{bitetrad}, that in Ref. \cite{Krssak:2017nlv} a bimetric approach was applied in the context of teleparallelism, where the connection-generating metric was assumed to be the Minkowski). In any case, bimetric variational principle deviates from the C-theory by relaxing the conformal constraint upon $\hat{g}_{\mu\nu}$. 
Thus, a different case can be considered, where one does not impose an {\it a priori} conformal relation, but instead regards the metric $\hat{g}_{\mu\nu}$ as a completely free dynamical variable
(relaxing also the symmetry requirement of the metric was considered \cite{BeltranJimenez:2012sz}). However, the linear $f(g^{\mu\nu}\hat{R}_{\mu\nu})=g^{\mu\nu}\hat{R}_{\mu\nu}$ theory, where both the $g_{\mu\nu}$ and the  $\hat{g}_{\mu\nu}$ are free dynamical variables, is not viable since one of the metrics then unavoidably becomes a ghost \cite{BeltranJimenez:2012sz,Golovnev:2014aca}. The action can be, by partial integrations, manipulated into a
form where there appear two metrics with identical kinetic terms, up to the crucial sign. In this formulation, it becomes more transparent to see how the fact that GR features
the metric in two physically and geometrically distinct roles, could be a result of peculiar kind of a higher symmetry. With the bimetric variational principle, 
is necessary to impose the relation of the metrics in order to avoid the ghost, and the case of GR is obtained with the choice $C=1$. 
This way, we at least heuristically understand the dynamics as the result will feature a double copy of metric kinematics. 

This feature is made explicit in our Palatini theory of STG. Geometrically, the covariant derivative of the gravitational connection is the Lie derivative. Therefore, in the resulting improved field theory, these covariant derivatives systematically copy the translational symmetries of the kinetic terms. 

\subsubsection{The kinematic factors in the gravity amplitudes}

In the computation of scattering amplitudes, the Feynman rules for the graviton $\delta g_{\mu\nu}$ 
depend upon the chosen coordinate system, but the dependence cancels out in the final scattering amplitude. 
The fact that the final result must be independent of the coordinate system is due to the diffeomorphism
invariance, and it is possible to understand this as a type of gauge symmetry acting
on the graviton field $\delta g_{\mu\nu}$. The Feynman rules are often considered in the De Donder gauge, 
$\nabla^\mu \delta g_{\mu\nu} = \frac{1}{2}\nabla_\nu \delta g$, and due to the abundance of terms generated at the vertices
the computations are much more onerous than in the usual Yang-Mills theory. However, in the final gauge-invariant
expressions for the amplitudes there typically occurs considerable cancelling of terms. The result could be much simpler than suggested at the
starting point of the traditional computation, and one may ''feel that a simple result ought to be obtained in a simple way'' \cite{Feynman:1996kb}. 

There is some evidence in string theory that such a short-cut could be found, and an interesting formula has been discovered that facilitates the computation of the gravity
amplitudes by relating them to the ''squares'' of gauge theory amplitudes. Kawai, Lewellen and Tye 
had derived a relation, which expresses any closed string tree
amplitude in any number of dimensions as a sum of products of two open string tree amplitudes \cite{Kawai:1985xq}. In
the field theory limit of the string amplitudes, this relation implies
that any graviton scattering amplitude can be expressed in terms of a sum of products of
colour-ordered gluon scattering amplitudes. More recently, the  work of Bern, Carrasco and Johansson established the 
{\it double-copy} structure in the gravity amplitudes \cite{Bern:2010ue}, which allows to obtain those amplitudes quite directly from
the Yang-Mills amplitudes by a curious ''squaring'' procedure.  Associated with each cubic diagram of the Yang-Mills amplitudes
is a colour factor and a kinematic numerator. Bern, Carrasco and Johansson had found that when a set of three colour factors satisfies a
Jacobi identity, then the kinematic numerators can be chosen such that they also satisfy the same Jacobi identity, and with this choice
of ''generalised gauge'' (involving some field redefinitions), the graviton scattering 
amplitude was shown to be obtained from the Yang-Mills amplitude by simply replacing the colour factors by the kinematic numerators.

As explained in Ref. \cite{Monteiro:2011pc}, ''the fact that these Jacobi identities are satisfied at all points strongly suggests
that there is a genuine infinite dimensional Lie algebra at work in the theory. Moreover, since
the gravitational amplitudes are formed by replacing Yang-Mills colour factors with these
kinematic numerators, this group seems to entirely determine the gravitational scattering
amplitudes.'' To uncover the hidden symmetry, first were explored the self-dual sectors of the gravity and the Yang-Mills theory.
The kinematic algebra in these sectors was made completely manifest by the comparison of the equations of motion of the two theories.  
The algebra was associated with diffeomorphisms. More precisely, it was written down as the Poisson brackets for the 
area-preserving diffeomorphisms in the two-dimensional space spanned by the Minkowski space light-cone coordinates. 
A correspondence was thus found between the Lie algebra of area-preserving
diffeomorphisms and the special unitary Lie algebra in its planar limit. 
In Ref. \cite{Monteiro:2011pc} it was concluded that ''in the case of gravity, the presence of an infinite dimensional algebra seems entirely
reasonable, and it is pleasing to see such an algebra play the role in gravity that the finite
dimensional colour algebra plays in Yang-Mills theory''  \cite{Monteiro:2011pc}. A new symmetry has been proposed for the Yang-Mills
side of the duality \cite{Brown:2016mrh}.

However, the origin of the symmetry in the gravity side remains undisclosed. It is somehow hidden in the canonical
approach of graviton perturbation theory, even when it is known that the double copy structure persists in classical field theory, where
some solutions of GR have been directly mapped to gauge theory backgrounds \cite{Monteiro:2014cda}.
 In Ref. \cite{Cheung:2016say}, it was recognised that the 
structure ''suggests a very surprising fact about the underlying symmetries of gravity'', and in an attempt to make the conjectured extra symmetry
manifest, GR was reformulated by using a double set of Lorentz indices, thus bestowing the theory a two-fold Lorentz invariance. 
However, it was not yet clarified how to perform the double copy construction for the scattering amplitudes in gauge theory and gravity in this approach. Also, the
previous result of Ref. \cite{Monteiro:2011pc} supports our intuition that the freedom to do the kinematical rearrangements has to do with a translational rather than a
(pseudo-)rotational invariance. As we have shown, there indeed is a double-diffeomorphism symmetry in GR, which becomes transparent in its $\mathcal{Q}$-formulation in STG.
It is remarkable that this simple formulation might finally elucidate the emergence of the spacetime geometry from the dynamics of closed strings, and the 
relationship between the fields in the gravity theory and those in the Yang-Mills theory that emerge from the open string dynamics.

\subsection{The Euclidean action}
\label{sec:euclidean}
We will conclude our discussion on interesting applications of the STG by computing the entropy of a Schwarzschild black hole within the usual formulation of GR and its two teleparallel equivalents in terms of torsion and non-metricity respectively. We will compute the entropy with the Euclidean action method. We will assume that the matter fields are absent. The thermodynamical properties of the black hole can be obtained from a grand partition function given by the path integral over the gravitational fields\footnote{We would consider also the integral over $\Gamma^\alpha{}_{\mu\nu}$ and the Lagrange multipliers, or the inertial
form of the $\Gamma^\alpha{}_{\mu\nu}$. For the purposes of the present discussion this will not be relevant.}
\begin{equation}
\mathcal{Z}=\int D[g]e^{-\mathcal{S}_{\rm E}(g)}
\end{equation}
with the Euclidean action $\mathcal{S}_{\rm E}(g)=-i\mathcal{S}(g)$, which is obtained by means of a Wick rotation $t=-i\tau$ so that $\tau$ has a period $\beta$ that gives the inverse of the black hole Hawking temperature. The dominant contributions to the partition function will come from the paths close to the classical field configuration. Let us assume $\mathcal{S}_{\rm E}(g)= \mathcal{S}_{\rm E}(\bar{g})+\mathcal{S}^{(1)}(\delta g)+\cdots$, where $g=\bar{g}+\delta g$. The logarithm of the partition function then becomes
\begin{equation}
\ln \mathcal{Z}=-\mathcal{S}_{\rm E}(\bar{g})+\ln \left( \int D[g]e^{-\mathcal{S}^{(1)}(\delta g)} \right)\,.
\end{equation}
We will further assume that the background contribution is the dominant contribution. Also, we will set $G=1$ in this subsection instead of $16\pi G=1$ as used in precedent sections. Once we have the on-shell Euclidean action, the entropy can be simply computed for a Schwarzschild solution with mass $M$ as
\be
S=\beta M+\ln\mathcal{Z}=\mS_{\rm E}
\ee
In GR, the computation of the Euclidean action of a Schwarzschild black hole is based on the Gibbons-Hawking-York (GHY) approach \cite{PhysRevD.15.2752}, where the standard Einstein-Hilbert term is supplemented with the GHY boundary term. This term is in fact necessary in order to have a well-defined variational principle without having to impose extra-conditions on the normal derivatives of the metric on the boundaries. Furthermore, a normalisation term determined by some fixed background geometry is included so that the Euclidean action vanishes for a Minkowski background. With all these elements, the total action reads
\begin{eqnarray}
\mathcal{S} &= & \mathcal{S}_{EH} +  \mathcal{S}_{GHY}+\mathcal{S}_{C} \nonumber\\ 
& = & \frac{1}{16\pi}\int_{\mathcal{M}}d^4x\sqrt{-g}\mathcal{R}  +  \frac{1}{8\pi}\int_{\partial\mathcal{M}}d^3y\epsilon\sqrt{|h|}(K-K_0)\,, \nonumber
\end{eqnarray}
where $K$ is the trace of the extrinsic curvature, $K_0$ is the extrinsic curvature of the background fixed geometry, $h$ the induced metric on the boundary with the coordinates $y$, and the sign $\epsilon$ depends upon whether the boundary is space-like or
time-like. The Euclidean Schwarzschild solution is given by
\begin{equation}
\diff s^2=\left( 1-\frac{2M}{r}\right)\diff\tau^2+\left( 1-\frac{2M}{r}\right)^{-1}\diff r^2+r^2\diff\Omega^2\,, \label{smetric}
\end{equation}
which represents an asymptotically flat spacetime with periodicity $\beta=8\pi M$ in Euclidean time. For this background configuration, the Einstein-Hilbert action vanishes $\mathcal{S}_{EH}=0$, hence the unique contribution arises from the GHY term
\begin{eqnarray}
\mathcal{S}_{\rm E} &=&\frac{1}{8\pi}\int_{0}^{\beta}\diff \tau\int_0^{2\pi}\diff\phi\int_0^\pi d\theta\sqrt{|h|}(K-{K}_0)\nonumber\\
&=&\frac{\beta}{2}M=\frac{\beta^2}{16\pi}=4\pi M^2\,, \label{gre}
\end{eqnarray}
where $\sqrt{|h|}=\left( 1-\frac{2M}{{r}_0}\right)^{1/2}{r}^2_0\sin\theta$, where ${r}_0$ is the radial coordinate at the boundary. We obtain here the well-known result for the entropy of a black hole in GR. Before proceeding to the teleparallel equivalents, let us stress here once again that the full Euclidean action is determined by the boundary term, which has been chosen here as the most extensively used GHY. This however presents an ambiguity caused by the fact that this boundary term is not unique if all we require is to have a well-defined variational principle without additional boundary conditions (i.e., on normal derivatives of the metric) without modifying Einstein equations (see for instance \cite{Charap:1982kn}). 

For comparison we shall perform the same computation in the teleparallel formulation of GR. For this it will be helpful to use the relation (\ref{relationRT}) which relates the torsion scalar of the teleparallel formulation with the Ricci scalar of conventional formulation of GR $-\mathring{\mathbb{T}} =\mathcal{R} +2\mathcal{D}_\mu T^\mu$ (which is equation (\ref{relationRT}) with $R=0$). Since for the Schwarzschild solution the Ricci scalar of conventional GR vanishes $\mathcal{R}=0$, we further have $-\mathring{\mathbb{T}} =2\mathcal{D}_\mu T^\mu$. Therefore, it does not matter whether we actually perform the integration over $-\mathring{\mathbb{T}}$ or $2\mathcal{D}_\mu T^\mu$. We can rewrite the teleparallel action, including the suitable counter-term, as
\begin{equation}
\mathcal{S}   =    \mathcal{S}_{\rm TEGR}  +   \mathcal{S}_{C} 
 = \frac{1}{8\pi}\int_{\partial\mathcal{M}}\diff^3y\epsilon\sqrt{|h|}n_\mu(T^\mu-{T}^\mu_0)\,, \label{bhtegr}
\end{equation}
where $n_\mu$ is the normal vector. 
Using the fact that $T^{\alpha}{}_{r\alpha}=\frac{2}{r}+\frac12\left( 1-\frac{2M}{r}\right)^{-1}\partial_r\left( 1-\frac{2M}{r}\right)$ and ${T}^{\alpha}{}_{r_0\alpha}=\frac{2}{r_0}$, one obtains for the Euclidean action
\begin{eqnarray}
\mathcal{S}_{\rm E}&=&\frac{1}{8\pi}\int_{0}^{\beta}d\tau\int_0^{2\pi}d\phi\int_0^\pi d\theta\sqrt{|h|}n_r(T^r-{T}^{r}_0)
\nonumber\\
&=&\frac{\beta}{2}M=\frac{\beta^2}{16\pi}=4\pi M^2\,, \label{tegre}
\end{eqnarray}
which agrees completely with the expression we obtained with the GHY approach. The fact that we do not need to include the GHY term separately in the teleparallel equivalent of GR was recently noted also 
 in \cite{Oshita:2017nhn}. Still, the counter term needs to be included to prevent the action from diverging. The counter-term in general is not unique, and its choice can be reflected in the physical results. 
As mentioned in the introduction, it is known that by finding the appropriate Lorentz frame in teleparallel gravity, one may renormalise the action without adding a separate counter term \cite{Lucas:2009nq,Krssak:2015rqa,Krssak:2015lba}. In the Palatini formulation of the theory, it becomes especially clear that we need not resort to the frames or the Lorentz connection, but instead may find spacetime coordinates such that the infra-red divergence of the action may be eliminated. If we transform the black hole metric from the Schwarzschild (\ref{smetric}) coordinates to some other coordinate system, the first term in (\ref{bhtegr}) stays
invariant, and the classical predictions are unaffected, but the divergent behaviour of the action may be tamed without the ${T}^\mu_0$-term.   

We will finalise our brief tour on the Euclidean action for the Schwarzschild black hole solution with the symmetric teleparallel formulation of GR. We will perform the computation directly in the coincident gauge where the original Diff symmetry is used to completely remove the connection. This however results in a Lagrangian (the Einstein or $\Gamma\Gamma$ formulation of GR) that only realises a Diff invariance up to a total derivative. While the equations of motion are oblivious to these total derivatives, the Euclidean action is sensitive to them as we have already seen in the precedent cases. This means that the value of the black hole entropy by resorting to the Euclidean integral will depend on the chosen coordinates where the black hole solution is expressed or, more correctly,  the coordinate basis that we use. Thus, in the coincident GR formulation we have replaced the ambiguity in the choice of the boundary term by an ambiguity in the choice of basis. 
The relation between the non-metricity scalar and the Ricci scalar of conventional GR is given by equation (\ref{ricciscalarq}) with $R=0$, thus $-\mathcal{Q}=\mathcal{R} +   \mathcal{D}_\alpha ( Q^\alpha - \tilde{Q}^\alpha )$. Furthermore, since $\mathcal{R}=0$ for the Schwarzshild solution, we simply have $\mathcal{Q} = -  \mathcal{D}_\alpha ( Q^\alpha - \tilde{Q}^\alpha ) = -\frac{2}{r^2}$.
For comparison, let us start by computing the Euclidean action directly in the spherical Schwarzschild coordinates given in (\ref{smetric}), which gives:
\ba
\mS_{\rm E}&=&\frac{1}{16\pi}\int\diff^4x\sqrt{-g}\mQ\nonumber\\
&=&-\frac{1}{16\pi}\int_0^\beta\diff\tau\int\diff\Omega\int_{2M}^{r_0} r^2\diff r\frac{2}{r^2}\nonumber\\
&=&\beta\left(M-\frac12r_0\right)=8\pi M\left(M-\frac12r_0\right).
\ea
We see here the same linear divergence in $r_ 0$ that is found in the classical computation in terms of curvatures and that is regularised by a normalisation boundary term. A similar regulator should be included here to obtain the physical result. Let us now explore in some detail how the value of the Euclidean action depends on the coordinate basis and, thus, an appropriate choice can lead to a finite result. 

For this purpose, we can transform (\ref{smetric}) to the isotropic coordinates so the Euclidean line element reads
\be
\diff s^2 =  \frac{\lp 1-\frac{M}{2R}\rp^2}{\lp 1+\frac{M}{2R}\rp^2}\diff \tau^2 
 + \lp 1+\frac{M}{2R}\rp^4 \diff \ell^2\,, \label{isotropic}
\ee
where we have defined $R=\sqrt{x^2+y^2+z^2}$ and $\diff \ell^2 = \diff x^2 +  \diff y^2 +  \diff z^2$. These coordinates relate to the usual Schwarzschild coordinates by
\be
r=R\left(1+\frac{M}{2R}\right)^2
\ee
so that the horizon in the Lorentzian metric is now located at $R=M/2$. In order to further illustrate that the differences appear when considering different coordinate basis, we can use the above isotropic coordinates in a spatial cartesian basis with coordinates $(x,t,z)$ or in a spatial spherical basis with coordinates $(R,\theta,\phi)$. In the spherical basis we obtain a divergent part that goes as $\mS_{\rm E}\sim Mr_0$ and, thus, we obtain the same type of divergence as in the precedent computations. However, in the cartesian basis, the Euclidean action is
\ba \label{stgraction}
\mS_{\rm E}&=&\frac{1}{16\pi}\int\diff^4x\sqrt{g}\mQ\\
&=&\frac{M^2}{8\pi}\int_0^\beta\diff\tau\int\diff x\diff y\diff z\frac{1}{R^4}\nonumber\\
&=&\frac{M^2}{8\pi}\int_0^\beta\diff\tau\int\diff\Omega\int_{M/2}^{R_0} R^2\diff R\frac{1}{R^4}\nonumber\\
&=&\frac{M^2}{2}\beta \left(\frac{2}{M}-\frac{1}{R_0}\right)\underset{R_0\to\infty}{\longrightarrow}\beta M=8\pi M^2.\nonumber
\ea
It is important to notice that the improvement in the convergence of the Euclidean action has been caused by the behaviour at the asymptotic infinity of the measure in the volume element $\sqrt{g}$ which, while in the spherical basis goes as $\sim R^2$, in the cartesian basis goes as $\sim1$. Let us stress that the $R^2$ factor in the above integration is due to the change to spherical coordinates in the cartesian basis (i.e., the Jacobian of the transformation). Thus, we see that in this basis the Euclidean action directly leads to a finite result without the need to consider boundary terms nor an appropriate normalisation. However, the entropy in this non-divergent coordinate basis would appear to be double the usual result. This merits further investigations\footnote{ An intriguing speculation is that the factor of 2 may be related to 
topology. In de Sitter space, the Bekenstein-Hawking entropy is quarter the horizon area, as in the conventional case of the black hole, but if the the geometry is constructed with real
projective spatial sections by identifying the antipodal points, the de Sitter entropy is one half the horizon area \cite{1126-6708-2003-09-009}. Though this was, for the physical reasons of avoiding
a kind of action-at-distance, already considered by  
Schwarzschild (and later de Sitter), the physical implications of identifying the antipodal points in their geometries is being currently taken under investigation \cite{Hooft:2016itl,Ong:2016vwr}.}. 

Here we have resorted to the computation of the Euclidean action in order to obtain the entropy of the black hole. In many cases, this coincides with the computation by means of Wald's formula \cite{Wald:1993nt}, which gives the entropy in terms of Noether charges and is valid for general theories with diffeomorphism invariance \cite{Wald:1993nt,Iyer:1994ys,Iyer:1995kg}. However, in the coincident GR, this symmetry is only realised up to a boundary term so it is not applicable\footnote{Of course the symmetry is restored by taking into account the connection. Then the most straightforward application of the Wald's formula would involve a the conjugate curvature \ref{density1}, which is given by the rank-four lagrange multiplier tensor density in STG. This would be the only case we are aware of that one actually needs to consider the solutions for the Lagrange multipliers in practice.}. It would be interesting to adapt a similar formalism for the class of theories treated here and elucidate its relation with the Euclidean action.

Note that we consider the metric (\ref{isotropic}) in the coincident gauge, i.e. still keeping the connection vanishing. Of course, did we transform both the connection and the metric, there would be
no difference either in the first line or the second line in (\ref{stgraction}). The example of this subsection illustrates this subtlety of the ''double-Diff'' symmetry, which allows to find the physical description of the system with both its {\it normalisation} and {\it renormalisation} (i.e. we are in the normal coordinates in the sense that our connection vanishes, plus our Euclidean actions are convergent).

\section{Conclusions}
\label{concsec}

This article presented Palatini theories of gravity in teleparallel geometry. Starting merely with the metric and the affine connection,
we have shown that it is possible to formulate teleparallel gravity without introducing a frame field or an additional set of indices associated to a tangent space by simply imposing the desired constraints with Lagrange multipliers. In the metric-compatible teleparallel torsion theories, the Lagrange multipliers that impose the constraints on the geometry are crucial for the dynamics of the theory, in contrast to the usual formulation in terms of frame (i.e. tetrad) fields. On the other hand, in the frame formulation of a symmetric teleparallel theory, one needs to solve for the Lagrange multipliers to obtain the full dynamics of the physical fields, but the Palatini formulation of a symmetric teleparallel theory is simpler since there the Lagrange multipliers decouple from the field equation of the metric. This  complementary structure of the two formulations of the two classes of geometries is briefly summarised in Table \ref{table}.  

In hindsight, there is a more straightforward way to analyse these geometries. In what was called the ''inertial variation'' (to distinguish from a handful of subtly different variational recipes considered in \cite{Golovnev:2017dox,Koivisto:2011vq}), one simply considers the pure-gauge
connection, instead of a general connection, in the action. Then no Lagrange multipliers are needed. Equivalently, if one first uses a Lagrange multiplier, solves the constraint
equation and plugs the solution for the connection back into the action, the inertial variation follows. This is implicitly done with the metric-compatibility constraint,
when one restricts the spin connection to be antisymmetric a priori, instead of starting in the metric-affine gauge theory and then constraining 
it to the context to Poincar\'e by a Lagrangian gauge fixing. So, the same can be also done explicitly with the teleparallelity constraint that sets the
curvature to zero, as considered in \cite{Golovnev:2017dox}. The general solution is then given by a $n\times n$ matrix that parameterises a general 
linear transformation. Of these only $n(n-1)/2$ remain in the metric-compatible case. In the less explored torsion-free i.e. symmetric teleparallel geometry, the pure-gauge connection is parameterised by four functions. The inertial variation is applicable in this geometry as well. 
 
\begin{center}
\begin{tabular}{ |c|c|c|c| }
\hline
$R^\alpha_{\phantom{\alpha}\beta\mu\nu}=0$ & Palatini formalism & Frame formalism \\
\hline
\multirow{3}{6em}{Teleparallel} & $Q_\alpha^{\phantom{\alpha}\mu\nu}=0$ & 6 inertial dof's \\ 
& dynamics from $\lambda$  & $\lambda$ decouples \\ 
& section \ref{telesec} & MAG \cite{Obukhov:2002tm} PGT \cite{Blagojevic:2000pi} \\ 
\hline
\multirow{3}{6em}{Symmetric teleparallel} &  $T^\alpha_{\phantom{\alpha}\mu\nu}=0$  & 4 inertial dof's \\ 
& $\lambda$ decouples & dynamics from $\lambda$ \\ 
& section \ref{symsec} &  \cite{Adak:2005cd,Adak:2006rx,Adak:2008gd} \\ 
\hline
\end{tabular} \label{table}
\captionof{table}{A summary of some properties of the two types of geometries in the two variational formalisms. 
When it comes to workings of the Lagrange multipliers, the structure of the theories is the opposite in the 
gauge variation (not discussed here) and in the Palatini variation.}
\end{center}

In this article we used quadratic theories as the example. Generalisations to nonlinear functions of the quadratic invariants is straightforward. 
Such generalised theories are interesting in modelling of dark energy, the origin of which is theoretically unknown but the properties of which can be experimentally tested. A plethora of gravitational dark energy models have been proposed previously, recall e.g. \cite{Cai:2015emx,Olmo:2011uz,Capozziello:2015lza,BeltranJimenez:2017doy}. The simplest
Palatini models run into problems because their lack of dynamics\footnote{They can be recasted as GR with modified matter sources, and then, once the dust source in cosmological background is made to accelerate the expansion
of the universe, the large-scale structure formation is in conflict with the observed one because the effective dark matter source, as seen by the modified gravity, is not cold but has pressures 
\cite{Koivisto:2006ie} (see, however \cite{Koivisto:2007sq}).}, which is not cured by simply adding torsion. However, in the teleparallel geometry, the torsion naturally acquires dynamics, and as shown
here, this can be realised in the Palatini formulation. We also showed that the STG promote the lapse function into a physically relevant variable.
This means that the time dilation rate should be taken into account besides the usual expansion rate, and the physical implications of qualitatively new feature in cosmology call for investigation.
 Amongst other interesting generalisations of the actions would be the inclusion of parity-odd sector and the boundary terms to facilitate the metric-connection reformulations of the self-dual theory \cite{Mielke:1992te}. Yet, one might consider teleparallel geometries where both torsion and non-metricity are present. Finally, when taking into account additional fields, these teleparallel geometries present a totally
 new framework for generalisations of the Horndeski theories. We can clearly reformulate all Riemannian Horndeski theories in teleparallel geometry, but there should exist, probably infinite classes of,
 second order field theories which are not admissible from Riemannian geometry. Some systematic studies are presently undertaken to explore scalar-torsion theories in teleparallelism
 \cite{Hohmann:2018rwf,Hohmann:2018vle,Hohmann:2018ijr,Hohmann:2018dqh}, and the first step into STG was just taken by introducing Brans-Dicke-type coupling of a scalar field and
 the $\mathcal{Q}$-scalar \cite{Jarv:2018bgs}.

The pure-gauge interpretation of the equivalence principle leads to the general linear symmetry of the affine connection being reduced to the diffeomorphic symmetry of the coordinate transformations. 
This suggests a new foundation for the gravitational geometry. The curvature of spacetime itself is not fundamental\footnote{As clarified in section \ref{spinor}, the Levi-Civita connection (\ref{christoffel}) and
and its familiar Riemann curvature nevertheless play their physical roles, since fermions are coupled as in GR. ''Matter geometry'' is metrical, though the ''gravitational geometry'' is trivialised.
Actually, this framework may turn out to vindicate the original intuition
of both Riemann and Clifford, according to which the spacetime curvature has to do with the presence of matter. }, but gravitation is rather described by the curvature of the internal geometry.
The canonical frame is now identified by the absence of affine spacetime curvature and the canonical coordinates are now identified by the absence of 
inertial effects. The tautological nature of this reasoning is corroborated by the two elementary findings we made in the STG: the connection is a pure translation, and there exists a unique quadratic form which decouples this inertial translation. This offers a physical rationale how to proceed with quantisation, where some of the technical benefits of the $\Gamma\Gamma$ action, whose covariant version
we effectively obtain, are already well-known \cite{Tomboulis:2017fim}. The unique claim of coincident GR is that there we can establish both the frame and the coordinate system wherein the canonical commutation relations can be recovered for the operators corresponding to physical observables. The {\it purified gravity} could be incorporated into the equations of quantum mechanics by considering them in a general frame, which presents us a totally new approach to arrive at the limit of quantum field theory \cite{tHooft:2002yjb} from a more robust conceptual foundation. According to the
equivalence principle, quantum mechanics in a non-inertial frame should be equivalent to quantum mechanics in the presence of a gravitational field, but we may only take advantage of this
after having first landed into the inertial frame in the coincident GR.

\acknowledgements{L.H. acknowledges financial support from Dr.
Max R\"ossler, the Walter Haefner Foundation and the
ETH Zurich Foundation. JBJ acknowledges the support of the Spanish MINECO's Centro de Excelencia Severo Ochoa Programme under grant SEV-2016-0597 and projects FIS2014-52837-P and FIS2016-78859-P (AEI/FEDER). TK acknowledges useful discussions with M. Kr\v{s}\v{s}\'ak and Y. C. Ong.}

\appendix 

\section{The metric propagator}

We consider the linear action for the metric and the connection (\ref{h_action}).
It can be shown \cite{Conroy:2017yln} that by integrating
out the inertial connection, the effective action for the metric which recovers the manifest Diff invariance, as it should. This is not manifest in (\ref{h_action}) alone since we have fixed the coincident
gauge, i.e. we have transformed the connection to zero. It might (or might not) be instructive to demonstrate explicitly how we recover 
Conroy's result \cite{Conroy:2017yln} and to extract the degrees of freedom that can be associated solely to the metric field, now 
working in the coincident gauge\footnote{In \cite{Conroy:2017yln} no gauge fixing was exploited and the theory was not restricted to second derivative order, the generic action then having double the number of terms in comparison to (\ref{abcd2}). In the more conventional notation \cite{VanNieuwenhuizen:1973fi,Conroy:2017yln}, the terms in (\ref{h_action}) would correspond to $a=c_1$, $b=\frac{1}{2}(c_2+c_4)$, $c=\frac{1}{2}c_5$, $d=c_3$, and the higher-derivative term discussed below would come with the coefficient $f$. For an
original analysis see \cite{VanNieuwenhuizen:1973fi}, and for a review of the pure metric theories including the $f$-term, see \cite{Biswas:2013kla}. The analysis of the Poincar\'e gauge theory (without higher derivatives) was completed only recently with odd-parity invariants \cite{Karananas:2014pxa}, and yet remains to 
extended to the metric-affine gauge theory.}. We first write down the important equation (\ref{cEoM}) in the first order in $h_{\mu\nu}$ in the coincident
gauge,
\ba
0 & = & c_1 \Box \partial^\nu h_{\nu\alpha}  +   \frac{1}{2}\lp c_2 + c_ 4\rp \lp \partial_\alpha\partial^\mu\partial^\nu h_{\mu\nu} + \Box\partial^\nu h_{\nu\alpha}\rp  \nn \\
 & + & c_3 \Box \partial_\alpha h +  \frac{1}{2}c_5\lp \Box \partial_\alpha h + \partial_\alpha\partial^\mu\partial^\nu h_{\mu\nu}\rp\,. \label{cEoMp}
\ea
Assuming that the TDiff condition is not identically satisfied with vanishing connection, we find the constraint equation for the divergence of the metric perturbation,
\be \label{notdiff}
\partial^\mu h_{\mu\alpha} = c_6 \Box^{-1}\partial_\alpha \partial^\mu\partial^\nu h_{\mu\nu} + c_7 \partial_\alpha h\,, 
\ee
where we defined the combinations of the parameters
\ba
c_6 & \equiv & -\frac{c_2+c_4+c_5}{2c_1+c_2+c_4}\,, \\
c_7 & \equiv & -\frac{2c_3+c_5}{2c_1+c_2+c_4}\,. 
\ea
The equation (\ref{notdiff}) allows us to rewrite the second term in the action (\ref{h_action}) as
\ba
\partial_\alpha h_{\mu\nu}\partial^\mu h^{\alpha\nu} & \rightarrow & c_6^2 \Box^{-1}(\partial^\mu\partial^\nu h_{\mu\nu})(\partial^\alpha\partial^\beta h_{\alpha\beta}) \nn \\
&  + & 2c_6c_7 \partial_\mu h^\mu{}_\nu\partial^\nu h + c_7^2 \partial_\alpha h \partial^\alpha h\,.  \label{2rw}
\ea
Further, we can take the divergence of the equation (\ref{cEoMp}) to obtain the formal solution for the trace of of the metric perturbation,
\be \label{traces}
h = -c_8\Box^{-1}\partial^\mu\partial^\nu h_{\mu\nu}\,,
\ee
where 
\be
c_8 = \frac{c_1+c_2+c_4+\frac{1}{2}c_5}{c_3+\frac{1}{2}c_5}\,.
\ee
Using this solution, we can rewrite the third and the fourth terms in the action (\ref{h_action}) as
\ba
\partial_\alpha h \partial^\alpha h  & \rightarrow & c_8^2 \Box^{-1}(\partial^\mu\partial^\nu h_{\mu\nu})(\partial^\alpha\partial^\beta h_{\alpha\beta})   \label{3rw} \\
\partial_\mu h^\mu{}_\nu\partial^\nu h & \rightarrow & -c_8 \Box^{-1}(\partial^\mu\partial^\nu h_{\mu\nu})(\partial^\alpha\partial^\beta h_{\alpha\beta}) \,, \label{4rw}
\ea
respectively. We could also use both (\ref{notdiff}) and (\ref{traces}) to obtain another expression for the fourth term in the action (\ref{h_action}) as
\ba
\partial_\mu h^\mu{}_\nu\partial^\nu h & \rightarrow & c_6c_8 (\partial^\mu\partial^\nu h_{\mu\nu})(\partial^\alpha\partial^\beta h_{\alpha\beta}) \\
& + & c_7 \partial_\alpha h \partial^\alpha h\,,
\ea
which is consistent with the previous expressions (\ref{2rw},\ref{3rw},\ref{4rw}). The higher-derivative term that is generated by the solutions to the constraint (\ref{cEoMp})
has to be taken into account in the action in order to obtain it in the Diff invariant form for all the constraints. It is well-known that with the four terms present in (\ref{h_action}), the only Diff-invariant combination is the Fierz-Pauli Lagrangian which describes only a spin-2 field, but by inclusion of the higher-derivative term the Lagrangian include also a propagating scalar mode without violating the Diff invariance \cite{Biswas:2013kla}. Collecting the results into the action, there remains only trivial algebra that yields us the final combination
\be
c_9 = \frac{c_3+\frac{1}{2}c_5}{c_1+c_2+c_3+c_4+c_5} - c_3\,,
\ee
which enters the result in the following way:
\ba \label{h_action2}
\mathcal{L} &\rightarrow &c_1\lp \partial_\alpha h_{\mu\nu} \partial^\alpha h^{\mu\nu} - 2\partial_\alpha h_{\mu\nu}\partial^\mu h^{\alpha\nu}\rp 
\nonumber\\
& + & c_9\lp 2\partial_\mu h^\mu{}_\nu\partial^\nu h - \partial_\alpha h \partial^\alpha h\rp \nn \\
& + & \lp c_1-c_9\rp  \Box^{-1}(\partial^\mu\partial^\nu h_{\mu\nu})(\partial^\alpha\partial^\beta h_{\alpha\beta})\,.
\ea
This is the expected form of the Lagrangian, which is familiar from the Riemannian context \cite{Biswas:2013kla}. It propagates an additional scalar field when $c_9 \neq c_1$, and results in the equations of motion that are equivalent to those Conroy recently derived in STG \cite{Conroy:2017yln}. 

\section{An electromagnetic analogy}
\label{sec:em}

In this subsection we briefly consider the much simpler theory of the spin-0 pure-gauge electromagnetism, to illustrate some features which 
are found to occur also in gravitation, as described in the spin-2 representation coupled to the pure-gauge connection.  

Let there be a photon $F_{\mu\nu}=2\partial_{[\mu}A_{\nu]}$ and a complex scalar $\phi$ charged under the $U(1)$ of the photon through the covariant derivative $\nabla_\alpha = \partial_\alpha - iA_\alpha$. In analogy with the teleparallel theories considered in the bulk of this paper, we will introduce a Lagrange multiplier to impose a vanishing $U(1)$ {\it curvature} so that the Lagrangian is given by
\be \label{u1}
\mL = -\frac{1}{4}F_{\mu\nu}F^{\mu\nu} + \lambda^{\mu\nu}F_{\mu\nu} + (\nabla_\alpha\phi)^\ast(\nabla^\alpha\phi)\,. 
\ee
The equations of motion for the independent components of the field are
\be \label{peoms}
\Box \phi = 0\,, \quad \Box \phi^\ast =0\,.
\ee
The equation of motion for the photon field $A_\mu$ has the form
\be
\partial_\nu\lp\lambda^{\mu\nu} + F^{\mu\nu}\rp = j^{\mu}\,,
\ee
with $j^\mu$ the Noether current associated to the global $U(1)$ symmetry of the $\phi$-sector. It determines or not the $\lambda^{\mu\nu}$, this $\lambda^{\mu\nu}$ decouples from the dynamics of other fields. Independently of the previous equation, the equations of motion for the fields (\ref{peoms}) imply 
that $\partial_\mu j^\mu=0$, where
\be \label{current}
j_{\mu} = i\lp \phi^{\ast}\nabla_\mu\phi - \phi^{\ast}\nabla_\mu\phi\rp\,.
\ee
For the classical purposes, we can integrate out the equation of motion for $\lambda^{\mu\nu}$, whose solution $A_\mu = \partial_{\mu}\psi$ so that we have the equivalent  two-scalar theory
\be \label{u2}
\mL_\flat = \alpha\lb\partial_\alpha-i(\partial_\alpha\psi^\ast)\rb\phi^\ast\lb\partial^\alpha+i(\partial^\alpha\psi)\rb\phi\,. 
\ee
By varying this Lagrangian wrt $\psi$, we also obtain that $\partial_\mu j^\mu=0$. This is the Golovnev's electromagnetic inertial variation, which allows the simplification of (\ref{u1}) to
(\ref{u2}). Further, one sees that the pure-gauge theory in the unitary frame is equivalent with the unconnected theory. By this we mean that the dynamics of the two-scalar theory (\ref{u2})
are equivalently described by the single scalar theory
\be \label{u3}
\mL_{1\flat} = \partial_\alpha\phi^\ast\partial^\alpha\phi\,,
\ee 
when one restricts to the specific gauge $\psi=0$. This is a consequence of the $U(1)$ symmetry of the action (\ref{u1}), and the same simplifications could be performed for 
any more complicated Horndeski $U(1)$-symmetric complex scalar theory. In the case of teleparallel gravity, Golovnev showed that generic actions with global Lorentz symmetry are 
''covariantised'' by the inclusion of the pure-gauge spin connection, though its equations of motion do not contain new information \cite{Golovnev:2017dox}. Therefore it is possible to consistently restrict oneself to the ''pure-tetrad'' formulation in teleparallel gravity, in analogy with the above reduction of (\ref{u1}) ''pure-scalar'' teleparallel electromagnetism (\ref{u3}).

These observations might clarify the logic of determining the canonical 1) geometry and 2) coordinate system. The equivalence principle of is the rationale for the transformations that determine 
1), and the principle of relativity of 2) specifies the action that is invariant under these transformations. By construction extending to the invariance under the gauge group of the desired presentation, the localisation of any theory should be subject to this logic. 
To reduce the theory of the line bundle into scalar theory the above condition 1) straightens the string and 2) synchronises the phases. In teleparallel gravity, the analogous requirements are 1) the flatness of the affine connection and 2) the independence of the spin connection. In the STG, the further implications are understood as  1) the symmetry affine connection and 2) the coincidence of the coordinate systems. Translations in spacetime are then understood as the integrable general linear transformations,  
in analogy with the phase transformations in the integrable electromagnetism. We then find that in the generic actions, the affine connection has an aetheric role of ensuring the conservation, 
even when this connection can be transformed to zero in the unitary gauge. Nevertheless, only the action that is oblivious to the affine connection has the claim of a global translation symmetry.
\newpage

\bibliography{Telepala}

\end{document}